\documentclass[superscriptaddress,twocolumn,floatfix,notitlepage,citeautoscript]{revtex4-1}
\setcitestyle{super}

\usepackage{upgreek}
\usepackage{siunitx}
\usepackage{bm}

\usepackage[normalem]{ulem}

\usepackage{fancyhdr}
\usepackage{amsmath}
\usepackage{graphicx}
\usepackage{color}
\usepackage[unicode=true,bookmarks=false,breaklinks=false,pdfborder={0 0 0},backref=false,colorlinks=true,allcolors=blue,anchorcolor=black]{hyperref}
\usepackage{breakurl}

\usepackage{lettrine}

\makeatletter

\@ifundefined{textcolor}{}
{
 \definecolor{BLACK}{gray}{0}
 \definecolor{WHITE}{gray}{1}
 \definecolor{RED}{rgb}{0.7,0,0}
 \definecolor{ORANGE}{rgb}{1,0.25,0}
 \definecolor{GREEN}{rgb}{0,1,0}
 \definecolor{BLUE}{rgb}{0,0,1}
 \definecolor{CYAN}{cmyk}{1,0,0,0}
 \definecolor{MAGENTA}{cmyk}{0,1,0,0}
 \definecolor{YELLOW}{cmyk}{0,0,1,0}
}

\newcommand{\changed}[1]{{\color{BLACK}#1}}

\newcommand{\nanotec}{
CNR NANOTEC, Istituto di Nanotecnologia, Via Monteroni, 73100 Lecce, Italy
}

\newcommand{\brisbane}{
ARC Centre of Excellence in Future Low-Energy Electronics Technologies, School of Mathematics and Physics, University of Queensland, St Lucia, Queensland 4072, Australia
}

\newcommand{\wolver}{
Faculty of Science and Engineering, University of Wolverhampton, Wulfruna Street, WV1 1LY, UK
}

\newcommand{\moscow}{
National Research Nuclear University MEPhI (Moscow Engineering Physics Institute), 115409 Moscow, Russia
}

\newcommand{\rqc}{
Russian Quantum Center, Skolkovo innovation city, 121205 Moscow, Russia
}

\newcommand{\iran}{
Department of Physics, Azarbaijan Shahid Madani University, Tabriz, Iran
}

\newcommand{\unisal}{
Dipartimento di Matematica e Fisica E.~de Giorgi, Universit\'{a} Del Salento, Campus Ecotekne, via Monteroni, Lecce, 73100, Italy
}

\newcommand{\infn}{
INFN sezione di Lecce, 73100 Lecce, Italy
}

\renewcommand\frontmatter@abstractwidth{\dimexpr\textwidth-0.5in\relax}
\makeatother

\begin{document}

\title{Full-Bloch beams and ultrafast Rabi-rotating vortices}

\author{Lorenzo Dominici}
\email{lorenzo.dominici@nanotec.cnr.it}
\affiliation{\nanotec}

\author{David Colas}
\affiliation{\brisbane}

\author{Antonio Gianfrate}
\affiliation{\nanotec}

\author{Amir Rahmani}
\affiliation{\iran}

\author{Vincenzo Ardizzone}
\affiliation{\nanotec}

\author{Dario Ballarini}
\affiliation{\nanotec}

\author{Milena De Giorgi}
\affiliation{\nanotec}

\author{Giuseppe Gigli}
\affiliation{\nanotec}
\affiliation{\unisal}

\author{Fabrice P.~Laussy}
\affiliation{\rqc}
\affiliation{\wolver}

\author{Daniele Sanvitto}
\email{dsanvitto@gmail.com}
\affiliation{\nanotec}
\affiliation{\infn}

\author{Nina Voronova}
\email{nsvoronova@mephi.ru}
\affiliation{\rqc}
\affiliation{\moscow}

\begin{abstract}
\textbf{
  Strongly-coupled quantum fields, such as multi-component atomic condensates, optical fields and polaritons, are remarkable systems where the simple dynamics of coupled oscillators can meet the intricate phenomenology of quantum fluids. When the coupling between the components is coherent, not only the particles number, but also their phase texture that maps the linear and angular momentum, can be exchanged. Here, \changed{on a system of exciton-polaritons, we have realized a so-called {\it full-Bloch beam}: a configuration in which all superpositions of the upper and the lower polariton---all quantum states of the associated Hilbert space---are simultaneously present at different points of the physical space, evolving in time according to} Rabi-oscillatory dynamics. As a result, the light emitted by the cavity displays a peculiar dynamics of \changed{spiraling} vortices endowed with oscillating linear and angular momentum and exhibiting ultrafast motion of their cores with striking accelerations to arbitrary speeds. This remarkable vortex motion is shown to result from distortions \changed{of the trajectories} by a homeomorphic mapping between the Rabi rotation of the full wavefunction on the Bloch sphere and Apollonian circles in the real space where the observation is made. Such full-Bloch beams offer new prospects at a fundamental level regarding their topological properties or in the interpretation of quantum mechanics, and the Rabi-rotating vortices they yield should lead to interesting applications such as ultrafast optical tweezers.}
\end{abstract}

\maketitle

\section{Introduction}

Some of the most counter-intuitive concepts of Physics arise from the representation that quantum mechanics brings to the usual notions of reality: one cannot refer to physical objects with definite properties and attributes, but only to measurements made on them. Despite the introduction of a wavefunction by the theory, descriptions most often remain in terms of localized objects: particles. Even when interactions between them are not weak perturbations, this interpretation in terms of objects is typically extremely robust and accurate, as one can preserve it by introducing new quasi-particles, such as electrons in the Fermi liquid of a metal, or bogolons in interacting Bose gases, or the so-called \emph{polaritons}, that arise when interactions between particles become too strong, as compared to the free energy, to be considered as a perturbation. Such conceptualizations are not reduced to the fundamental cases of interacting modes, but can also be applied to a wide family of topological defects. Solitons, for instance, that are nonlinear wave phenomena bearing all the characteristics of physical objects, can be better and accurately described as such.  Quantized vortices are another stunning example. They are the fundamental modes of rotation for fields mapped by a complex wavefunction (in particular, atomic Bose-Einstein condensates~\cite{Matthews1999}, superfluids~\cite{leggett_superfluidity_1999}, superconductors~\cite{blatter_vortices_1994}, electron
beams~\cite{grillo_holographic_2015,uchida_generation_2010}, and light~\cite{yao_orbital_2011,molina-terriza_twisted_2007,zhang_particlelike_2019}).
Since the wavefunction must reconnect with itself, its phase $\varphi$ has to undergo an integer number of twists when looping around the centre of rotation. This number, the topological charge, defines the (intrinsic) orbital angular momentum (iOAM) per particle, quantized in units of $2\pi$.
While vortices stem from a delocalized rotation in the entire space, in the hydrodynamic (long-wavelength) limit, they are neatly pictured as objects. This manifests most strikingly through the so-called \emph{vortex core}, a null-density point-like phase singularity in which the wavevector $\bm{k}$ diverges (since $\bm{k} = \nabla\varphi$).  Within optical fields, the connection between the Maxwell description and quantum dynamics is revealed by the so-called singular beams~\cite{willner_different_2012,franke-arnold_advances_2008}. Such vortex beams are most attractive due to their classical-to-quantum connections~\cite{review2019}, with easy experimental generation by the Laguerre-Gaussian laser modes that allow to combine iOAM within a beam with spin angular momentum (SAM)   related to photon polarization. Propagating light carrying angular momentum can be used for information processing~\cite{nagali_optimal_2009,leach_quantum_2010}, for instance in offering a new degree of freedom for multiplexing~\cite{Wang2012,gibson_free-space_2004}. Recently, curved light beams~\cite{Zhao2015} and standing or moving solenoidal beams~\cite{Lee2010} have been proposed as the most advanced tools for optical tweezers~\cite{padgett_tweezers_2011,Rahman2015}. The rich vortex-related phenomenology illustrates the importance of the recent advances with optical vortices, giving rise to the field of ``complex light'' or ``structured light''~\cite{secor_complex_2016}.

Among quantum fluids able to sustain vortices, microcavity exciton-polaritons~\cite{kavokin_book17a,Byrnes2014} have brought forward unique assets to study these objects~\cite{Rubo2007,Lagoudakis2009,
  roumpos_single_2011,Voronova2012,Dall2014,kartashov_rotating_2019,dominici_vortex_2015}, including their multi-component nature with coherent coupling and their 
Rabi oscillations~\cite{voronova_oscillations,rahmani_polaritonic_2016}. In the strong coupling regime of the two bare components---the cavity photon and the exciton fields---new normal modes of the system appear. The two eigenmodes are known as the upper (UP) and lower polariton (LP) modes, whose dispersions are separated in energy by the Rabi splitting. The polariton Rabi oscillations are known since the first
polariton studies, but their sub-picosecond imaging and control have only recently been achieved~\cite{Dominici2014,colas_polarization_2015}. While the vorticity transfer and dynamics in coherently coupled atomic BECs has been considered  theoretically~\cite{calderaro_vortex_2017,tylutki_confinement_2016}, here, based on the recent progress in both the quality of samples and the level of control and detection, we experimentally study for the
first time the joint polariton Rabi-oscillatory and spatio-temporal vortex dynamics. It yields a spectacular phenomenology illustrating how notions of `physical objects' must be treated with care and linked to the underlying full-wavefunction picture.
Namely, in the light emitted by the system, we observe a vortex, being at all times accurately defined and visible, undergoing striking ultrafast rotations with large accelerations and decelerations. This is achieved with a fine control over the shape of the fields formed in the microcavity and based on the coherent transfer of particles and their momentum between the exciton and photon modes, by preparing a wavefunction that realizes simultaneously \emph{all} the possible quantum states of the
polaritonic Hilbert space. Since this is, in our case, the simplest possible case of a two-dimensional Hilbert space---a Bloch sphere---we refer to this wavefunction as a \emph{full-Bloch beam} \cite{schultz_creating_2016}. The consequent photoluminescence (PL), {\it i.e.}, the measurement made solely on one of the Bloch vector projections (photons), reveals light structured in both space and time that can carry simultaneously all the three different kinds of angular momentum of an optical beam~\cite{Bliokh2015}: the created states are circularly polarized, which corresponds
to SAM, they carry the topological charge of a vortex (iOAM) initially imprinted by the excitation, and possess an extrinsic orbital angular momentum (eOAM) brought forward from the displacement of the vortex
core from the origin. Finally, since the centre-of-mass of such states as well as the vortex core inside and the net transverse linear momentum (nTLM) are also rotating in time, this endows them with time varying OAM~\cite{rego_generation_2019}.
We identify the appropriate mathematical description for such a richly-structured object, namely, we provide a conformal and bijective link between the real plane and the Bloch sphere that shows how the   complex dynamics in real-space of both the wavefunction and even more so of the vortex core, reduces to a trivial Rabi rotation on the Bloch sphere. Within the scope of this study, both the physical system and
its theoretical description remain at the most fundamental level without complex effects of many-body interactions or nonlinearities. In particular, we demonstrate how linear dissipation further adds interesting twists to an already puzzling dynamics, where the complex evolution can be represented in terms of a general M\"obius transform. We anticipate that increasing interactions in the OAM-carrying polariton systems~\cite{assman_2020} or combining various Bloch sphere rotations by delocalization in reciprocal space would result in the distortion of the full-Bloch beam metric, which is subject to further investigation.

The paper is organised as follows. In the next Section, Sec. II, we describe the experiments where a particular combination of the topology of excitation with the Rabi-oscillatory dynamics gives rise to the peculiar state and its dynamics mentioned above, and provide its reproduction by the theoretical model. We discuss the underlying concept that makes this complex phenomenology a beautiful example of a quantum mechanical projective measurement performed on the full wavefunction. In Sec. III, we show how, while the full wavefunction is
rigidly rotating with uniform speed on the Bloch sphere, its projection onto an ``observable'' state leads to amazing behaviors, including ultrafast motion up to superluminal speeds
and the periodically changing angular momentum without applied forces.
In the concluding Sec. IV, we discuss the concrete consequences and impact of our findings on polaritonics in particular and on the research fields that deal with dressed states of light-matter interactions and richly structured fields, in general.

\section{The Full-Bloch beam concept}

\begin{figure*}[htbp]
  \centering
  \includegraphics[width=1.00\linewidth]{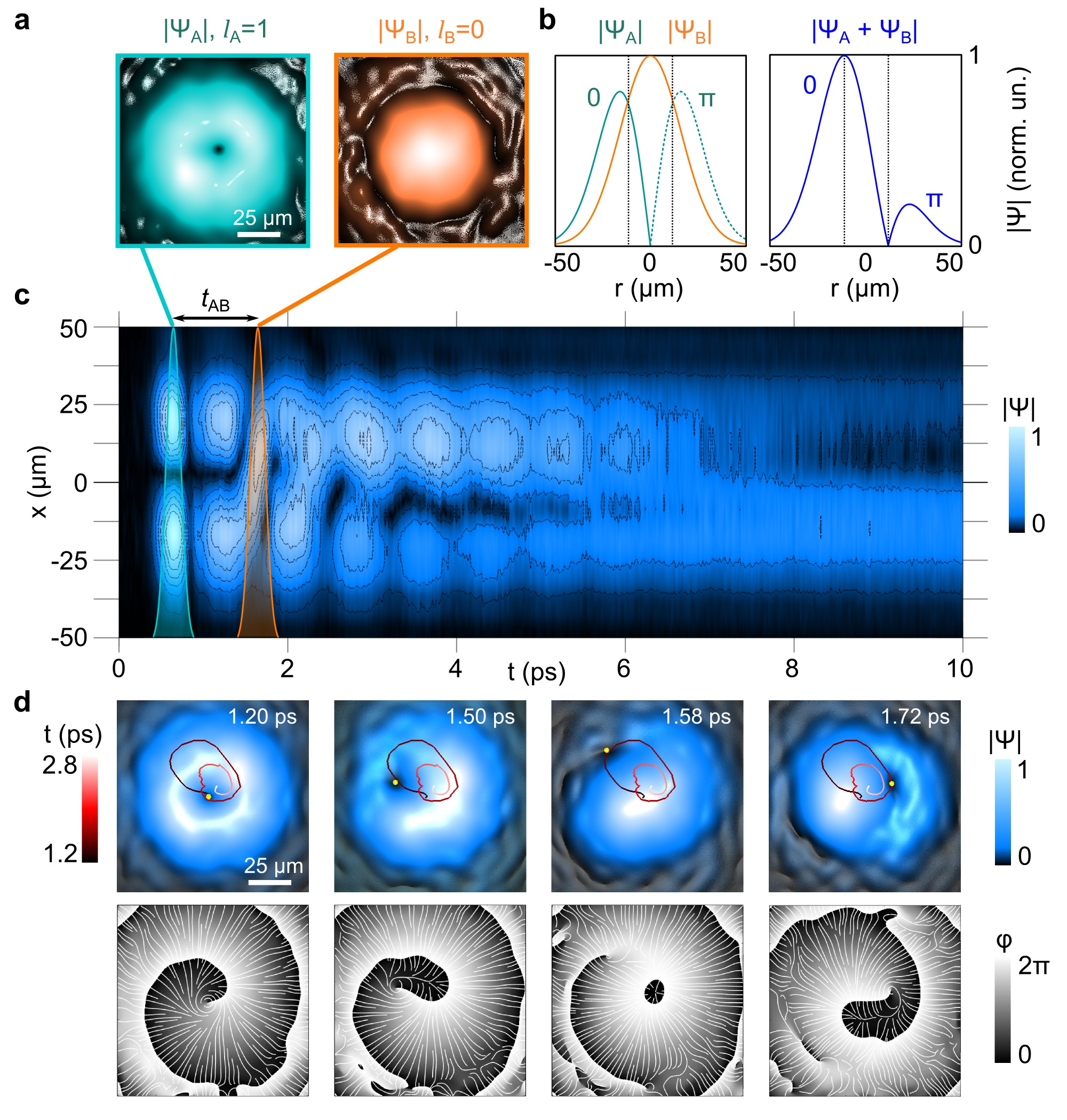}
  \linespread{1.0} \protect\protect\caption{\textbf{Rabi-rotating vortex experiment.}
  \textbf{a}, Photonic emission from the polariton \changed{field} excited by independent resonant pulses: the vortex $A$ and the plain Gaussian $B$.
  \textbf{b}, Displacement of the vortex core position by coherent overlap of the beams. The solid and dashed parts of the green line (radial cut of pulse $A$) are in anti-phase leading to new positions for both the maximum and null density ({\it i.e.}, vortex core) of the total beam at a radial distance set by constructive and destructive interferences.
  \textbf{c}, Timespace charts of the polariton amplitude along a central crosscut when superimposing the excitation and control pulses (time delay $t_{AB} \approx 1\text{ ps}$) leading to desynchronization of the Rabi oscillations along the diameter as shown by their bending after the arrival of pulse $B$.
  \textbf{d}, Time evolution of the polariton amplitude and phase at times $t = 1.20, 1.50, 1.58$, and $1.72~\text{ps}$. The vortex core, identified by the phase singularity, is marked by a yellow dot in the amplitude maps with the red curve showing its trajectory in a timespan of two Rabi periods ($t = 1.2 - 2.8~\text{ps}$). The white streamlines in the phase maps show the in-plane momentum vector field $\boldsymbol{k}_C=\nabla\varphi$. See also the Supplementary Movie~SM1~\cite{SMmovies}.
  }
\label{FIG_exp_rartex_maps}
\end{figure*}

\subsection{Initialisation and spiraling vortex dynamics}

The initialisation of the dynamics is performed by spatially overlapping two pulses generated externally to the sample, with and without optical vortices, delayed in time, with a fine control on the sought shape of the wavepacket formed in the microcavity.  Namely, we overlap a femtosecond Laguerre-Gauss $LG_{00}$ laser pulse onto an $LG_{01}$ pulse created by means of a patterned liquid crystal retarder (a
$q$-plate)~\cite{dominici_vortex_2015,Cardano2013}. The first excitation pulse (pulse $A$) directly generates the polariton \changed{field}, setting an initial condition to be a vortex state upon resonant photon-to-polariton conversion on the sample. The control pulse (pulse $B$), which is vortex-free, is obtained from the unconverted part out of the $q$-plate. The spatial structures of pulses $A$ and $B$ are shown in Fig.~\ref{FIG_exp_rartex_maps}a. Such density plots are the photonic emission maps from the polariton \changed{field} after an independent excitation by one of the two pulses. For simplicity, the pulses $A$ and $B$ are co-polarized (and chosen to be circular). In a further extension of this work, one could also realize a full Bloch-Poincar\'e beam by using different polarizations. The relative power between the two pulses $P_A/P_B$ is set close to~2, in order to have a comparable top density for the associated polariton populations at the time of the control pulse arrival. The temporal delay $t_{AB}$ between the two pulses is controlled by the difference in the two arms lengths (excitation delay line), defining the optical phase shift between the pulses $\varphi_{AB}$ (modulo $2\pi$). Finally, the PL from the cavity is monitored by the use of the digital holography technique~\cite{Dominici2014}, based on the time-resolved (detection delay line) homodyne interference that allows us to track in time with ultrafast precision the complex-valued photon field dynamics, {\it i.e.}, both its phase and intensity.

In the absence of the Rabi coupling, the result would be immediate~\cite{Satyajit_PRApplied}: pulse $B$, the plain-Gaussian $LG_{00}$ with a central top intensity and homogeneous phase, coherently sums up to the previous vortex field $LG_{01}$, which is empty in its centre. The central phase singularity is thus instantaneously displaced to a new position, by the destructive interference between the initial and the newly-generated field, as sketched in Fig.~\ref{FIG_exp_rartex_maps}b. In 2D, the specific direction of the displacement is set by the optical phase delay, \textit{i.e}., the new azimuthal position of the vortex core is set by the phase shift $\varphi_{AB}$ between the two pulses, which is one example of a feature which we are able to control coherently with high accuracy.  Similarly, a new density maximum is obtained in the radially opposite direction to the core. In presence of the Rabi coupling, however, since the normal modes of the system differ from the bare cavity mode, additionally to the displacement of the vortex, the obtained composite polariton \changed{field} is set into a very peculiar motion, with a rotation of the vortex core itself, as shown in Fig.~\ref{FIG_exp_rartex_maps}c and d (cf.~Supplementary Movie SM1~\cite{SMmovies} for a better illustration of the effect). Figure~\ref{FIG_exp_rartex_maps}c displays the chart of the photonic amplitude along a central crosscut, sampled with a $\delta t=\SI{20}{\femto\second}$ timestep. This chart makes evident the peculiar feature achieved by the use of a double excitation pulse with different topologies: the density imbalance along the crosscut, the filling of the empty core of the initial vortex by the second pulse, and, importantly, the desynchronization of the Rabi oscillations between the radially opposed regions. More strikingly, Fig.~\ref{FIG_exp_rartex_maps}d shows the photonic amplitude and phase maps of the \changed{field} at various times following the initial condition set, first, by the circular symmetric $LG_{01}$ pulse~$A$ carrying a vortex (at $t = \SI{1.20}{\pico\second}$) which is suddenly displaced aside upon the arrival of the second pulse $B$ ($t = \SI{1.50}{\pico\second}$). This triggers a complex dynamics of the vortex core inside the polariton spot, grazing its boundary ($t = \SI{1.58}{\pico\second}$) at ultrafast speed before coming back again near the spot centre with a huge deceleration ($t = \SI{1.72}{\pico\second}$) to start another cycle.  The solid line in the amplitude maps shows the vortex core trajectory in the timespan of two Rabi periods ($t =1.2$--$\SI{2.8}{\pico\second}$). The radial damping is due to the faster UP dephasing time
$\tau_{U}\sim \SI{2}{\pico\second}$~\cite{Dominici2014}. It can be roughly described as a spiraling orbit, where the full loop is travelled in a time of $T_R = \SI{0.78}{\pico\second}$, but with a varying speed along the trajectory (here $T_R=2\pi/\Omega_R$ denotes the Rabi oscillations period).
Note that the location of the vortex core is easily and unambiguously tracked in the phase mapping. A similar effect, with a smaller orbit, is obtained for the photon density centre-of-mass.

\begin{figure*}[htbp]
  \centering
  \includegraphics[width=1.00\linewidth]{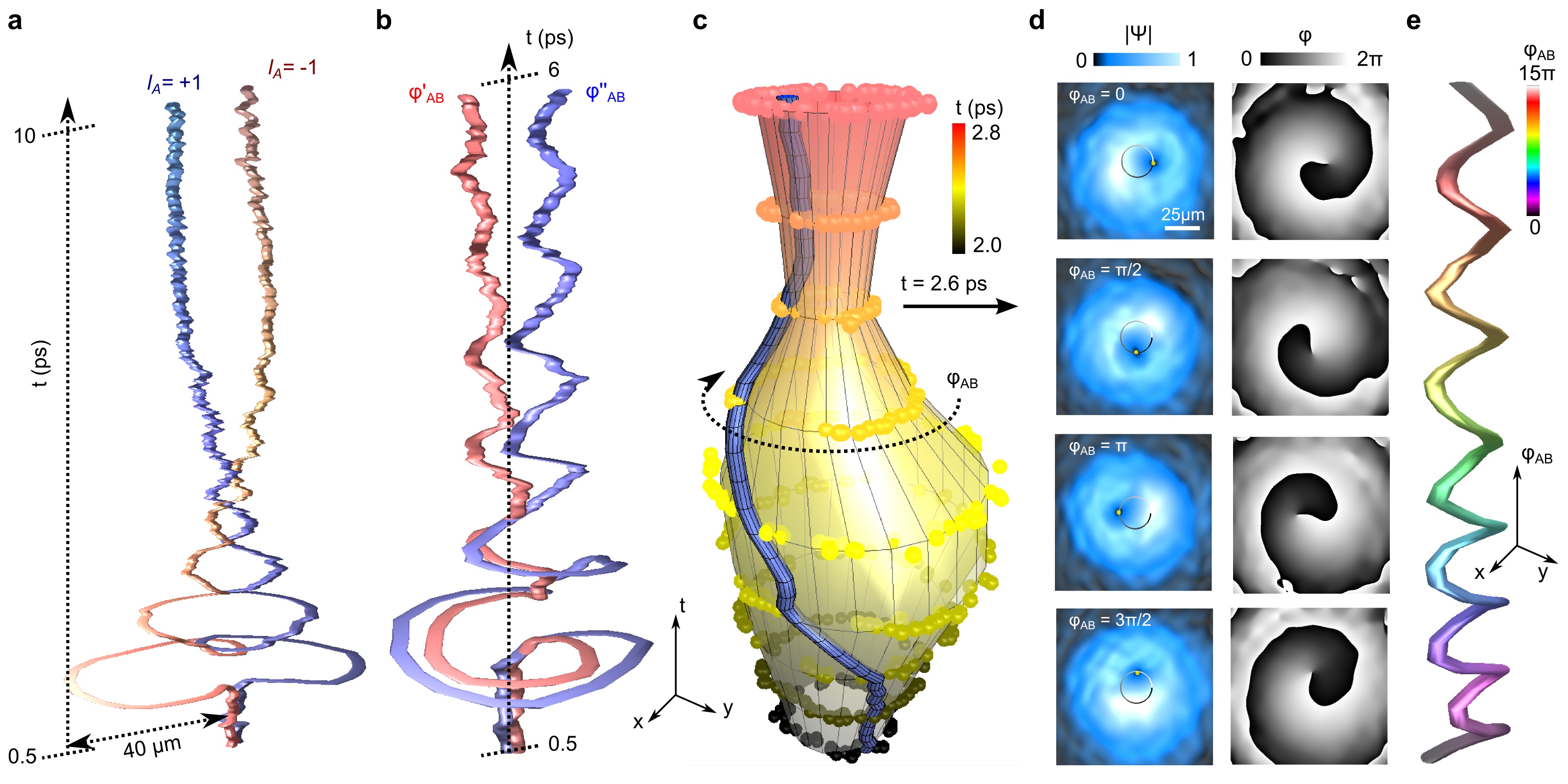}
  \linespread{1.0} \protect\protect\caption{\textbf{Coherent shaping of trajectories.}
  \textbf{a}, Experimental $xyt$ photonic vortex lines (time range $t=0.5-10\text{ ps}$, step
    $\delta t=0.02\text{ ps}$). The two realizations are for opposite $l_A=+1$ (blue) and $l_A=-1$
    (red) initial topological charges, resulting in opposite spiraling directions.
    \textbf{b}, Experimental $xyt$ vortex lines (time range $t=0.5-6.0\text{ ps}$) for two optical phase shifts $\varphi_{AB}$ between the excitation pulse $A$ ($LG_{01}$) and the control pulse $B$ ($LG_{00}$).
    \textbf{c}, Detail of the dynamics over a single Rabi period ($t = 2.0$--$2.8$~ps) with the $xyt$ surface topology bottle-envelope drawn by the different vortex lines when sweeping $\varphi_{AB}$. Its radius breathes with the Rabi oscillations. It is mapped by spheres at 100~fs time intervals, spanning $\varphi_{AB}$ in a 15$\pi$ range by successive $\sim\pi/4$ steps. The blue tube shows a 
    $xyt$ vortex line for a given $\varphi_{AB}$.
    \textbf{d}, Amplitude and phase of the polariton \changed{field} at a fixed time of the dynamics ($t = 2.6$~ps), for different realizations corresponding to optical phase shifts separated by $\pi/2$.  The vortex core, shown with a yellow dot in the amplitude maps, and the black/white circle of isotime positions from the previous panel.
    \textbf{e}, Helicoidal vortex string in the virtual $xy\varphi_{AB}$ space at a fixed
    time. }
\label{FIG_exp_bottle1}
\end{figure*}

\subsection{Fine control of the vortex lines}

The trajectory of the observed vortex core is reported as $xyt$ vortex lines in Fig.~\ref{FIG_exp_bottle1}a, in the $0.5$--$\SI{10}{\pico\second}$ time range, where the two vortex lines
correspond to two different realizations, upon starting with opposite iOAM ({\it i.e.}, opposite initial vortices imprinted by the pulse $A$, either $l_A=+1$ or $l_A=-1$). When starting with opposite windings, also opposite spiraling directions (eOAM) are obtained. Additionally, the fine delay control allows us to rotate the whole $xyt$ core trajectory around the $t$ axis, by changing the optical phase shift $\varphi_{AB}$ between the two pulses. As an example, the two specific realizations corresponding to slightly different values $\varphi_{AB}^\prime$ and $\varphi_{AB}^{\prime\prime}$ are shown in Fig.~\ref{FIG_exp_bottle1}b, in the $0.5$--$\SI{6.0}{\pico\second}$ time range.  The entire set of core trajectories describes a closed $xyt$ topological surface, upon scanning the phase shift $\varphi_{AB}$ along one complete $2\pi$ turn, as shown in Fig.~\ref{FIG_exp_bottle1}c.  The surface has been mapped in a limited time range (a single Rabi period, $2.0$--$\SI{2.8}{\pico\second}$) upon registering the phase singularity position (solid spheres) at egular time intervals of $\SI{0.1}{\pico\second}$, and by changing $\varphi_{AB}$ by successive $\sim\pi/4$ steps. Here the detail of a specific vortex string (blue tube sampled every $\SI{20}{\femto\second}$) demonstrates how the line is perfectly carved on the topological surface. It is observed that every $xy$ encircling path that encompasses the topological surface produces a unitary phase winding, for every time into the dynamics and every value of $\varphi_{AB}$. The closed surface is fundamentally symmetric around the initial vortex position, and its radius breathes with the Rabi period. The circular symmetry of the surface is a direct consequence of the symmetry of both the initial vortex pulse $A$ and the Gaussian pulse $B$, and of their coaxial alignment. The circular symmetry is further evident in the panels of Fig.~\ref{FIG_exp_bottle1}d, which report the amplitude and phase of the polariton \changed{field} at a fixed time of the dynamics ($t=\SI{2.6}{\pico\second}$) but for different phase delays ($\varphi_{AB}$ spaced by $\pi/2$ intervals, which in other terms is $\lambda/4$ length delay differences).  The vortex core is at this time displaced from the centre by always the same distance, but at a changing azimuthal direction that directly maps the optical phase shift, and describes an almost perfect circle (see black/white line in the density maps Fig.~\ref{FIG_exp_bottle1}d). The vortex line at a fixed time can also be represented in the $xy\varphi_{AB}$ space, where it assumes the helicoidal shape, as represented in Fig.~\ref{FIG_exp_bottle1}e.

\subsection{Vortex trajectory modelisation}\label{sec_models}

The observed core dynamics that arises from the interplay of coherent coupling and vorticity can be reproduced with the standard coupled Schr\"odinger equations (cSEs) that accommodate these two ingredients. The cSEs are written for the coupled exciton~$\psi_\mathrm{X}(x,y,t)$ and photon~$\psi_\mathrm{C}(x,y,t)$ macroscopic wavefunctions (see Appendix);
hence, while only the photonic field $\psi_\mathrm{C}$ is easily accessible in the experiments, the theory allows us to study as well the excitonic component $\psi_\mathrm{X}$, and the LP(UP) fields $\psi_\mathrm{L,U}= (\psi_\mathrm{C}\pm\psi_\mathrm{X})/\sqrt{2}$ (in the simplest case of zero detuning), or any other combination of those.

\begin{figure*}[htbp]
  \centering
  \includegraphics[width=0.80\linewidth]{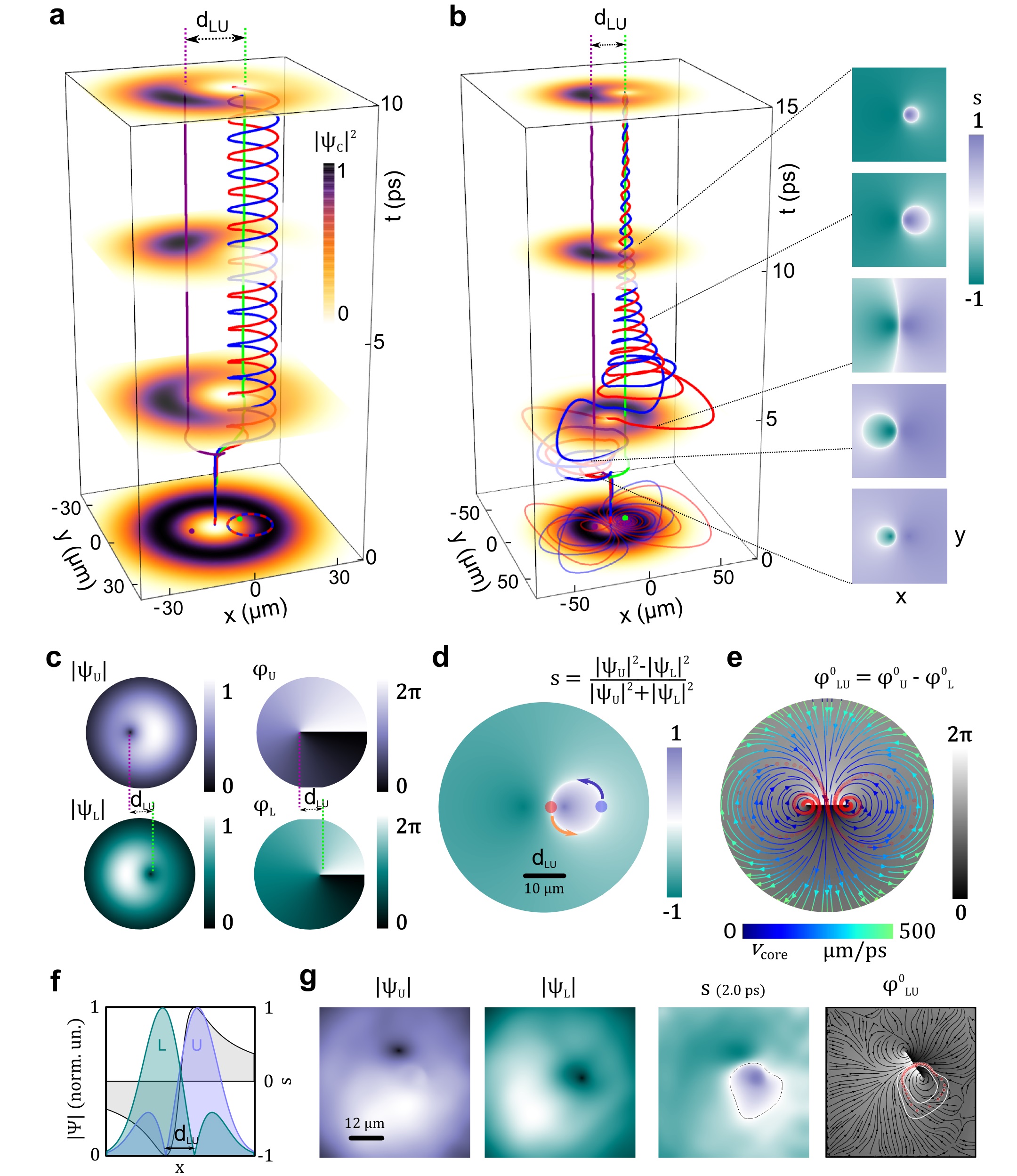}
  \linespread{1.0} \protect\protect\caption{
    \textbf{Theoretical trajectories and experimental reconstruction of the polariton profiles}. \textbf{a},\textbf{b},
    Numerical $xyt$ vortex lines (purple for the UP mode, green/LP, red/photon, blue/exciton). Photonic density $|\psi_\mathrm{C}(x,y)|^2$ maps are superimposed at selected time
    frames. Pulse $A$ provides coinciding vortex cores in all of the fields. Upon arrival of pulse $B$, the four cores are displaced to different positions. \textbf{a}, no decay,
    and exciting close to the LP. \textbf{b}, with UP decay rate $\gamma_U = 0.2$~ps$^{-1}$ and exciting close to the UP. The photon and exciton cores first circle around the UP line, while the global UP-LP relative content $S$ changes, resulting in the orbit radius growth, and then switch to rotation around the LP vortex core, finally shrinking as the UP population vanishes. The slow change of the instantaneous orbits in time due to the UP dephasing is shown in the local polariton imbalance parameter $s =
    (|\psi_\mathrm{U}|^2-|\psi_\mathrm{L}|^2)/(|\psi_\mathrm{U}|^2 +|\psi_\mathrm{L}|^2)$ maps on the right, where the white line is the isodensity $s=0$ at each chosen moment of time (a
    circle with drifting centre and  radius).
    \textbf{c}, Maps of the amplitudes and phases of the normal modes (polaritons), showing their
      density asymmetry and the displacement of the UP and LP vortices.
    \textbf{d}, Spatial distribution of the local polariton imbalance $s(x,y)$ in the case of
      higher LP content ({\it i.e.}, $S < 0$), as in \textbf{a}. The photon and exciton cores (red and blue dots) move along the UP-LP isodensity line $s = 0$.
    \textbf{e}, Spatial distribution of the relative phase $\varphi_{LU}^0$ for a vortex dipole. The photon core (red dots) follows the direction of the gradient $\nabla \varphi_{LU}$ (shown by streamlines whose color shows the core velocities, larger at the boundary:
    ${v}_\textrm{core} = \Omega_R /|\bm{\nabla}\varphi_{LU}|$).
    \textbf{f}, Side view of~{\bf c} showing the non-coinciding profiles
    in the LP and UP fields due to the vortices splitting by the distance $d_\textrm{LU}$. The right axis shows the changing local imbalance $s$ across the cut, for the case of equal global UP/LP populations ($S=0$).
    \textbf{g}, Left to right: experimental counterparts to the theoretical panels {\bf c}, {\bf
      d} and {\bf e}, obtained by fitting the measured photon density at each point $(x,y)$ and reconstructing the polariton fields as explained in the Appendix. The
      rightmost panel additionally shows the two instantaneous isodensity lines $s=0$ at the
      start (2.0~ps) and end (2.8~ps) of the second Rabi period, as well as the positions of the core (red dots) during this loop over the $\nabla \varphi_{LU}$ streamlines.}
\label{FIG_theo_basic}
\end{figure*}

Assuming at first the particle-conserving model with no decay terms, we show in Fig.~\ref{FIG_theo_basic}a the position of the vortex cores in both~$\psi_\mathrm{C}$ (red helic), corresponding to the one measured experimentally, and $\psi_\mathrm{X}$ (blue helic) as functions of time, along with the vortex cores in LP/UP (in green/purple lines), all of which are easily computed as the null-density points (phase singularities) of the respective wavefunctions. The excitation pulse $A$ creates a preset condition with the vortex perfectly centred into the beam, and coincident in all four fields. Upon arrival of the control pulse $B$, the $\psi_\mathrm{U}$ and $\psi_\mathrm{L}$ vortices split into fixed positions out of the centre of the beam, being separated by the distance $d_\mathrm{LU}$, as shown in Fig.~\ref{FIG_theo_basic}a,~b, and~c. As described above, the direction of the instantaneous displacement of the vortex core is defined by the time delay between the pulses $t_{AB}$. Here, since the normal modes of the system have different well-defined eigenenergies $\hbar\omega_\mathrm{U,L}$, the respective optical phase shifts $\varphi_{AB}$ between the two pulses for the UP and LP fields differ:
$\varphi_{AB}^\mathrm{U,L} = 2\pi(t_{AB} ~ \text{mod} ~ t_{opt}^\mathrm{U,L})/t_{opt}^\mathrm{U,L}$, where $t_{opt}^\mathrm{U,L}=2\pi/\omega_\mathrm{U,L}$. Thus the angular difference between the directions of displacement of the UP and LP vortices $\varphi_{AB}^\mathrm{U}-\varphi_{AB}^\mathrm{L}$ becomes defined by the coarse time-delay between the two pulses (of the order of the Rabi period $T_R$, in the ps scale) and hence by the associated Rabi phase $\Phi_{AB}= 2\pi(t_{AB} ~ \text{mod} ~ T_{R})/T_{R}$. The
UP-LP cores displacement directions are opposite and hence $d_\mathrm{LU}$ is maximal (as shown in Fig.~\ref{FIG_theo_basic}a, b, and~c) when the pulse $B$ is sent in the so-called anti-Rabi-phase (\textit{i.e.}, $t_{AB}$ is an half-integer multiple of $T_R$); on the contrary, if $t_{AB}$ is an integer multiple of $T_R$ (the two pulses are in Rabi-phase), the $\psi_\mathrm{U,L}$ vortices are displaced to the same point in space. The two leftmost panels in Fig.~\ref{FIG_theo_basic}g show the UP and LP cores displacement for the experimental realization of Fig.~\ref{FIG_exp_velocity}, where the UP and LP density profiles were reproduced from the measured photon wavefunction using the time-fitting by the theoretical model of the experimental data at each position (see Appendix for details of the experimental data fitting, and for additional modelisation cases of various time delays).

\subsection{Homeomorphism between real space and the Bloch sphere}

The displacement of the UP and LP vortices to different positions results in the non-coinciding spatial profiles of the two eigenstates $\psi_\mathrm{U,L}= |\psi_\mathrm{U,L}| e^{i\varphi_{U,L}}$ (as schematically represented in Fig.~\ref{FIG_theo_basic}f) and, as a consequence, in an instantaneous change of both the ratio $|\psi_\mathrm{U}|/|\psi_\mathrm{L}|$ and phase difference $\varphi_{LU}=\varphi_U-\varphi_L$ at each point in space. Effectively, it means that the polariton \changed{field} at each position $(x,y)$ is now characterized by its own relative content and relative phase between the upper and the lower polariton, changing from one point to the other. As can be seen in the upper part of Fig.~\ref{FIG_theo_basic}a, at the arrival of the second pulse, the vortex cores in the coherently coupled bare fields $\psi_\mathrm{C,X}=(\psi_\mathrm{U}\pm\psi_\mathrm{L})/\sqrt{2}$ start to
describe circular orbits powered by the Rabi oscillations. At every instant, the exciton and photon cores are on the $\pi$-opposite relative phase positions along the orbit, thus acting similarly to a Newton's cradle, with one core slowing down while the other one accelerates. The varying speeds of the cores are well visible in the vortex lines (see also the Supplementary Movies SM1 and SM2~\cite{SMmovies}): despite being perfect circles when projected on the $(x,y)$--plane, the $\psi_\mathrm{C,X}$ vortices cores orbits in Fig.~\ref{FIG_theo_basic}a are not perfectly helical in $xyt$ space due to their non-uniform speed.

\begin{figure*}[htbp]
  \centering
  \includegraphics[width=1.0\linewidth]{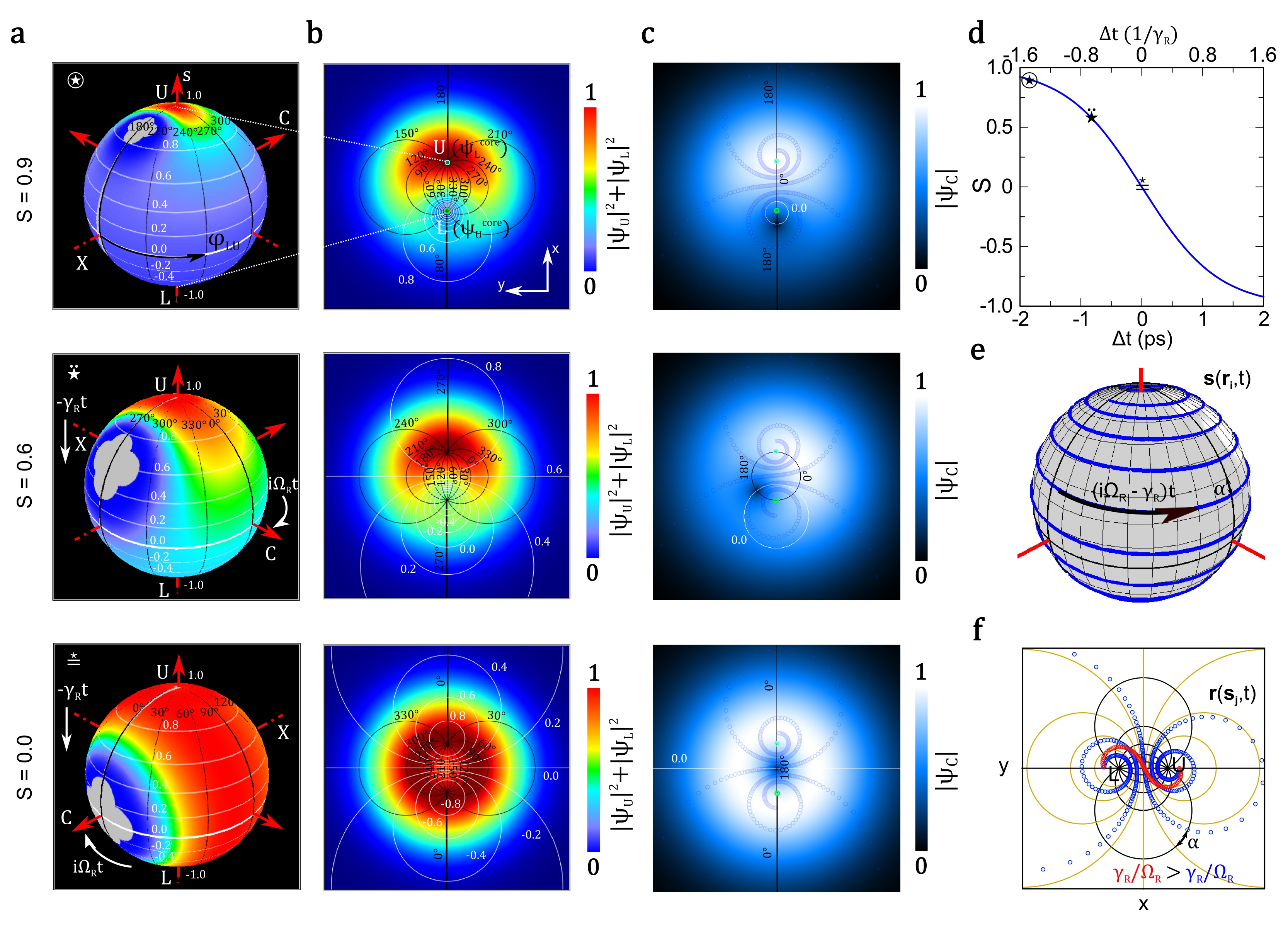}
  \linespread{1.0} \protect\protect
  \caption{\textbf{Full-Bloch beam on the sphere and its projection in real-space.}
  \textbf{a}, Total density $|\psi_\mathrm{total}|^2=|\psi_\mathrm{U}|^2+|\psi_\mathrm{L}|^2$
    covering the full Bloch sphere for three different values of $S=0.9$ (top), $0.6$ (middle) and $0.0$ (bottom). In this space, the dynamics of the full-wavefunction is a simple rotation ($i\Omega_R t$) around the vertical axis. With the differential decay the density $|\psi_\mathrm{total}|^2$ drifts downwards the L pole and the three panels shown correspond then to three successive times separated by $1.25 T_R$.
    \textbf{b}, Stereographic projection of the full-wavefunction from the Bloch sphere to the real space $|\psi_\mathrm{total}(x,y)|^2$ corresponding to the observed intricate dynamics.
    The two families of mutually orthogonal Apollonian circles (white and black) map the loci of
    equal polaritonic balance $s$ and isophase $\varphi_{LU}$. Their intersection identifies a single point on the Bloch sphere and thus a unique quantum state. See the Supplementary
    Movies SM3 and SM4~\cite{SMmovies} for better visualisation.
    \textbf{c}, Corresponding photon wavefunctions $\psi_\mathrm{C}(x,y)$, obtained by the projection~(\ref{projection}) of the full wavefunction. The white circles (in the bottom panel, of infinite radius) correspond to $s=0$ and thus to the trajectory of the photon vortex core in space, while its intersection with the isophase $\varphi_{LU}=\pi$ specifies the exact position of the core at any given time. The superimposed double spiral shows the photon vortex trajectory with decay, starting from $S=0.9$. For {\bf a}--{\bf c}, $\gamma_R/\Omega_R=0.1$.
    \textbf{d}, Evolution of the UP-LP global imbalance $S(t)$ due to decay. The curve can be started at any point, depending on the initial condition $S(0)$ defined by the excitation and is given here as a function of a time delay $\Delta t$ relative to $S=0$. The same curve is also followed locally at each point of the real space by the parameter $s(\bm{r},t)$, starting from the value given by its initial distribution $s(\bm{r}_i,0)$. The three symbols on the line correspond to the three times selected for the previous columns, as marked.
    \textbf{e}, Bloch-sphere trajectory $\bm{s}(\bm{r}_i,t)$ (blue curve) of a QS in presence of decay, for a given point $\bm{r}_i$ in real space, for $\gamma_R/\Omega_R=0.05$. The vector $\bm{s}$ defining the QS on the sphere combines the two spherical coordinates $(s,\varphi_{LU})$.
    \textbf{f}, Trajectories $\bm{r}(\bm{s}_i,t)$ (blue and red dots for the decay rates       $\gamma_R/\Omega_R=0.1$ [case of the previous columns] and~$0.5$, respectively) of a specific QS in real space, on top of the Bloch conformal metric which gets projected as two families of circles: yellow circles for the isodensities (a total of 9 orbits, $S$ from -0.8 to 0.8 with $\delta S = 0.2$, plus the UP and LP cores at $S = \pm 1$) and black circles for the isophases (12 circles with $\delta\varphi_{LU} = \pi/6$).
    With decay, the centres of the black circles drift and their radii change with time (see also the Supplementary Movie SM4~\cite{SMmovies}). The rhumb angle~$\alpha$ with respect to the parallels (constant~$s$) is defined by $\tan\alpha = \gamma_R/\Omega_R$ both on the sphere (\textbf{e})
    and in real space (\textbf{f}). Note that the panels \textbf{b} are rotated by $90^\circ$ compared to \textbf{f}, for a clearer correspondence with the Bloch spheres shown in \textbf{a}.
  }
\label{FIG_L2_model}
\end{figure*}

Interestingly, there is a privileged geometry to describe the Rabi-rotating vortex phenomenon, and this is not the 2D plane (in real or reciprocal, {\it i.e.}, Fourier-transformed, space) where a continuous
mapping of multiple wavefunctions ($\psi_\mathrm{C,X,L,U}$) should be done at each time. Instead, the best way to describe the system's intricate dynamics is on the polariton Bloch sphere~\cite{Dominici2014}. Each quantum state (QS) of the system can be expressed in the polariton basis on this sphere as follows:
\begin{equation}\label{qs}
|QS\rangle=\frac{1}{\sqrt{2}}\bigl(\sqrt{1-s}\,|L\rangle + \sqrt{1+s}\,e^{i\varphi_{LU}}\,|U\rangle\bigr),
\end{equation}
where $|L\rangle$ and $|U\rangle$, being the pure LP and UP states, correspond to the south and north poles of the sphere, respectively, and~$(s,\varphi_{LU})$ are the coordinates that pinpoint any of the possible quantum states of the system, \changed{with $s = (|\psi_\mathrm{U}|^2-|\psi_\mathrm{L}|^2)/(|\psi_\mathrm{U}|^2 + |\psi_\mathrm{L}|^2)$}
representing the {\it local} (position-dependent) UP-LP population imbalance and $\varphi_{LU}(x,y)$ their {\it local} relative phase. On the Bloch sphere, since the value of $s$ effectively defines the latitude of a quantum state (being exactly the projection on the vertical LP-UP axis), it can be seen as
an analogue to the Stokes parameter $S_3$ on the Poincar\'{e} sphere of polarizations. Normalized to the total density of excitations~$\mathcal{N}$, Eq.~(\ref{qs}) provides an absolute (space-independent) definition of the full-wavefunction. A choice of   space defines the parameters, for instance, in the physical 2D-plane, the normalization becomes \changed{$\mathcal{N}=\iint(|\psi_\mathrm{L}(x,y)|^2+|\psi_\mathrm{U}(x,y)|^2)\,dxdy$. Thus,} each QS is uniquely identified by a point \changed{(in any space) corresponding to} the coordinates $(s,\varphi_{LU})$ with the density~\changed{$|\psi_\mathrm{total}|^2=|\psi_\mathrm{L}|^2+|\psi_\mathrm{U}|^2$}.
The full wavefunction mapped onto the Bloch sphere is shown in the upper panel of Fig.~\ref{FIG_L2_model}a.  When the bare energies of the photon and exciton states are equal (no detuning), the Rabi-oscillatory dynamics corresponds to the rigid rotation of the Bloch sphere at a uniform speed around the vertical axis (see the three rows of Fig.~\ref{FIG_L2_model}a), so that in the absence of decay, the parameter $s$ stays fixed in time, and the temporal evolution of each QS is governed by the azimuthal angle changing continuously as $\varphi_{LU} = \varphi^0_{LU}(\bm{r}) + \Omega_R t$, {\it i.e.} the sum of the geometrical and dynamical phases, respectively. As discussed above, the spatial component of the relative phase $\varphi^0_{LU}(\bm{r})$ is not uniform in space, but has the shape of an UP-LP vortex dipole defined by the UP and LP spatial profiles, which is shown in Fig.~\ref{FIG_theo_basic}e (model) and~g (experiment). The spatio-temporal evolution $\varphi_{LU}(\textbf{r},t)$ is provided in the Supplementary Movie SM3~\cite{SMmovies}. Neglecting the spread in momentum on the dispersion relation, which is a good approximation for broad-enough wavepackets over the Rabi timescale, all states follow the same equator-parallel dynamics ({\it i.e.}, the QS trajectories are defined by the equation $s(x,y)=\text{const}$, see Fig.~\ref{FIG_L2_model}a and the corresponding lines in the real space in Fig.~\ref{FIG_L2_model}b). This treatment allows to easily track the trajectories of the photon and exciton vortices cores by looking for the trajectories where $s(x,y)=0$, since the points of zero-density in the respective fields $\psi_\mathrm{C,X}$ are bound to move along the lines where $|\psi_\mathrm{U}|^2 - |\psi_\mathrm{L}|^2 = 0$, which corresponds to the equator of the Bloch sphere (see Fig.~\ref{FIG_theo_basic}d showing the distribution of $s$ in space, and the line marked as $s=0.0$ in Fig.~\ref{FIG_L2_model}c that shows the photon wavefunction observed in the experiment). Since at each point of space the system assumes a different quantum state, it is no longer simply or even well described by the pairs $\psi_\mathrm{C,X}$ or $\psi_\mathrm{L,U}$ alone, and one needs instead to turn   to the full wavefunction of the system which is defined both as a function of the real plane $(x,y)$ and the \changed{Bloch sphere} $(s,\varphi_{LU})$ \changed{coordinates}.

The nature of the state which is actually created in the experiment by superimposing two pulses and displacing the vortices in the normal modes of the system, is a \emph{full-Bloch beam}, in the sense that \emph{all} quantum states of the Bloch sphere (all values $-1\leq s\leq1$ and $0\leq\varphi_{LU}<2\pi$),
except for just one specific state at each moment of time (see below), are simultaneously present somewhere in the physical space. Since each point in space realizes a different QS, the measurement of the ``photon field'' is made on the position-dependent photonic fraction of the polariton \changed{field}, hence the measured photon wavefunction is given by $\psi_\mathrm{C}(x,y,t)=|\psi_\mathrm{total}(x,y)|\langle   x,y|\Psi_C^\textrm{pr}\rangle$, where the projection of the normalized full wavefunction onto the photon state is given by
\begin{equation}\label{projection}
|\Psi_C^\textrm{pr}\rangle = |C\rangle \langle C|QS\rangle=\frac{1}{2}\bigl(\sqrt{1-s} + \sqrt{1+s}\,e^{i\varphi_{LU}}\bigr)|C\rangle\,.
\end{equation}
A similar expression can be obtained for the ``exciton field'' by projecting the normalized full wavefunction on the exciton state: $|X\rangle \langle X|QS\rangle$. From this perspective, the observed
vortex core corresponds to the QS of the polariton Bloch sphere, which, when projected onto a photon state according to Eq.~(\ref{projection}), provides a null-density, {\it i.e.}, $|\Psi_C^\textrm{pr}\rangle=0$, whereas $|\psi_\mathrm{total}|$ stays non-zero (while the neighbouring points create a phase singularity). Clearly, the photon vortex core corresponds to the $|X\rangle$ state on the sphere: by definition, this state has no photon component and is therefore the only point where $|\Psi_C^\textrm{pr}\rangle$ vanishes. Any other QS has a nonzero photon content. Similarly, the exciton vortex core corresponds to the point where $|QS\rangle=|C\rangle$. In fact, the topology of the full wavefunction is such that any point in real space becomes a vortex core in a certain basis of observables.

In the linear regime, \changed{the density of quantum states is preserved} on the Rabi-rotating sphere, and this must be true in space as well. As a consequence, any QS, for example the pure exciton state $|X\rangle$ (the $\psi_\textrm{C}$ vortex core), which is to be found at a given location at any given time, must drift continuously to another location in space in the next instant of time, with the same intensity. Consequently, the excitation of all the quantum states differing at all points of space creates a bicontinuous mapping, {\it i.e.}, a homeomorphism between the 2D Bloch sphere of quantum states and the 2D real-space physical plane (including infinity), as shown in Fig.~\ref{FIG_L2_model}a,~b. The metric consisting of parallels and meridians on the sphere is conformally ({\it i.e.}, angle preserving) mapped to the metric made of two mutually orthogonal circle families, or bipolar circular coordinates  (Apollonian circles) in real space (see in Fig.~\ref{FIG_L2_model}b, where the constant $s$ and $\varphi_{LU}$ are marked by the white and black lines, respectively). This physical realization of a mathematical homeomorphism between the extended complex plane (here: $(x,y)$--plane plus a point at infinity) and the Riemann sphere (here: the polariton Bloch sphere) results in a stereographic projection that maps circles on the sphere to circular trajectories on the plane, except the projection point associated with the QS that is at infinity in real space. Due to the Rabi rotation, this QS is not a fixed point on the sphere. The projection plane (the $(x,y)$--plane shown in Fig.~\ref{FIG_L2_model}b and~c) is at all times tangent to the sphere on the opposite side from the projection point. In all panels of Fig.~\ref{FIG_L2_model}a, the projection point on the Bloch sphere is the QS exactly in the middle of the grey area. \changed{This is the state mapped to infinity, and the Bloch vector tends to the same limiting value regardless of how the limit $\bm{r}\to\infty$ is taken on the plane.} At the same moment of time, the QS which is $\pi$-opposite to it on the sphere (on the same parallel) is projected onto the real space exactly in the middle of the UP-LP vortex dipole in the relative phase profile $\varphi_{LU}(x,y;t)$. This effect is best illustrated in the bottom row of the panels a--c in Fig.~\ref{FIG_L2_model}: when the projection point on the Bloch sphere coincides with the state $|C\rangle$, the $\pi$-opposite state ({\it i.e.}, $|X\rangle$) corresponds to the photon vortex core which is seen in (c) exactly in the middle between the stereoprojections of the points $|U\rangle$ and $|L\rangle$. The position of the projection point and hence the exact shape of the  stereographic projection from the Bloch sphere to the plane are defined by the {\it global} imbalance of the UP and LP populations in the system, namely, $S=\int(|\psi_\mathrm{U}|^2 - |\psi_\mathrm{L}|^2)   d\textbf{r} /\mathcal{N}$. The three rows in Fig.~\ref{FIG_L2_model}a,~b, and~c, corresponding to three different values of $S$, show snapshots of the full-wavefunction's total density  $|\psi_\mathrm{total}|^2$ on the Bloch sphere as well as its stereographic projection $|\psi_\mathrm{total}(x,y)|^2$ in real space and the corresponding profile of the photon wavefunction $|\psi_\mathrm{C}(x,y)|$ obtained by the QS-projection according to Eq.~(\ref{projection}), respectively. Only in the case of equal total populations $S=0$ (bottom row), the parallels $s=\text{const}$ are projected symmetrically to the $(x,y)$--plane (see the schematic representation of this case in Fig.~\ref{FIG_theo_basic}f). The equator of the sphere and hence the trajectory of the $\psi_\mathrm{C,X}$ vortex cores in this case is projected to a straight line going to infinity exactly in the middle of the UP-LP vortex dipole, symmetrically dividing the trajectories corresponding to parallels from the upper and the lower semi-spheres on the $(x,y)$--plane (see the bottom panels in
Fig.~\ref{FIG_L2_model}b and~c). However, when the system is excited closer to the UP energy ($0<S<1$), the projections of all parallels (the isodensity lines) shift toward the opposite semi-space, forming circles around the $\psi_\mathrm{U}$ core (point `L' on the real plane), as shown in the top two rows of Fig.~\ref{FIG_L2_model}b, and around the LP core (point `U') when excited closer to the LP energy ($-1<S<0$). Examples of cSEs dynamics in different cases, depending on the excitation energy, are presented in Fig.~\ref{FIG_theo_basic}a,~b (together with the maps of $s(x,y)$), and in the Appendix.

Through the homeomorphic mapping, the evolution of any QS can be described as transformations of the extended complex plane, belonging to the M\"{o}bius group\cite{needham}: $f(z)=(az+b)/(cz+d)$. This can be reduced to the simple Rabi rotation by associating the complex number $z$ with the QS coordinates on the Bloch sphere as $z=\psi_\mathrm{U}/\psi_\mathrm{L} = \sqrt{(1+s)/(1-s)} \exp(i\varphi_{LU})$, which yields $a=\exp(i\Omega_Rt)$, $b=c=0$, and $d=1$ (alternatively, this can be seen as a rigid rotation of the total density on a motionless sphere, see Fig.~\ref{FIG_L2_model}a). The two fixed points of the M\"obius transformation in real space are then represented by the positions of the LP and UP vortex cores, which correspond to the $|U\rangle$ and $|L\rangle$ states on the Bloch sphere, respectively (see the points marked as `U' and `L' in the corresponding top panels of Fig.~\ref{FIG_L2_model}a and~b). These are the only two points which cannot undergo any change, {\it i.e.}, being eigenstates, or normal modes, they preserve themselves.

\subsection{Differential decay and the drifting orbits}

To describe the full dynamics including the evolution of the global polariton imbalance parameter $S$, we include the UP and LP mode decays $\gamma_{U,L}$, respectively, to the cSEs model (see Appendix). While there would be little to understand in the modifications of the dynamics due to decay in the real-plane alone, the connection to the Bloch sphere through the homeomorphic mapping unveils a full family of elegant geometrical constructs that explain the observations from the evolution of the full-wavefunction on the Bloch sphere towards the LP pole in the presence of dissipation. Figure \ref{FIG_theo_basic}b shows the numerical dynamics of the $\psi_\textrm{C,X,L,U}$ cores obtained for
$\gamma_U = \SI{0.2}{\pico\second}^{-1}\gg\gamma_L$, with the excitation and control pulses whose energy is set close to the UP mode, while the three rows of Fig.~\ref{FIG_L2_model}a--c show the different QS trajectories on the plane and on the sphere (the isodensity circular orbits) for the early times of the dynamics at the three successive times corresponding to $S=0.9$, $S=0.6$, and $S=0$ (top to bottom). The local imbalance map $s(x,y)$ and the isodensity circular orbits change in time due to the reshaping of the total density on the Bloch sphere with the change of $S(t)$, and the trajectory of the two cores $s(x,y)=0$ (corresponding to a local equality $|\psi_\mathrm{U}|^2 = |\psi_\mathrm{L}|^2$) can be seen as a ``domain wall'' between the areas of excess of the UP and LP densities in real space ($s>0$ and $s<0$, the upper and the lower semi-spheres on the Bloch sphere), as shown in the snapshots on the right-hand side in Fig.~\ref{FIG_theo_basic}b, as well as on the relevant experimental snapshot in Fig.~\ref{FIG_theo_basic}g. As one can see, the $\psi_\mathrm{C,X}$ vortex cores first rotate around the $\psi_\mathrm{U}$ core, with a growing orbit radius. When the UP population decreases to the level of the LP mode, the orbit becomes a straight line between the positions of the $\psi_\mathrm{U,L}$ cores (corresponding to the limit of a circle with infinite radius). The initial configuration hence reverts, and the $\psi_\mathrm{C,X}$ vortex cores switch to spiraling from clockwise to anticlockwise direction, around the $\psi_\mathrm{L}$ core now, and with the decaying orbital radius as $S$ keeps changing. Finally, they both overlap into the position of the LP core, as this is the surviving mode at long times. For a better visualisation, we refer to the Supplementary Movie SM3~\cite{SMmovies}. The actual vortex trajectories hence result from the orbiting along these expanding/shrinking circles, with a drifting centre.  Their $xy$--projection is a double spiral which is shown in Fig.~\ref{FIG_theo_basic}e and Fig.~\ref{FIG_L2_model}c.
After the arrival of the pulse $B$ and the initiation of the dynamics, the difference in dephasing
$\gamma_R=\gamma_U - \gamma_L\approx \gamma_U$ results in the global imbalance $S$ changing with time as
$S(t) = \tanh\{-\gamma_Rt +\text{atanh}[S(0)]\}$, which is shown in Fig.~\ref{FIG_L2_model}d. This hyperbolic tangent curve can be started at any point $S(0)$ that is set by tuning the initial excitation energy between $\hbar\omega_\textrm{U}$ and $\hbar\omega_\textrm{L}$. The decay alone, resulting in the decrease of $S(t)$ with time, can be seen as the motion of the total density $|\psi_\mathrm{total}|^2$ downwards along the surface of the Bloch sphere, as represented in the successive panels (top to bottom) of Fig.~\ref{FIG_L2_model}a. Similarly to the rotation of the sphere, which belongs to the {\it elliptic} subclass of the M\"{o}bius transformations, the dynamics produced by different decay rates of the two normal modes can be seen as the dilation of the sphere, thus corresponding to the {\it hyperbolic} subclass of the M\"{o}bius group\cite{needham}. Like the elliptic transformation, the hyperbolic transformation fixes two points (in our physical case, the points `U' and `L', {\it i.e.}, the positions of the lower polariton and upper polariton vortex cores, respectively, in real space), however the corresponding trajectories of the QS on the plane would create a family of circular arcs, along which the points flow away from the first fixed point (U) and towards the second one (L). When united together, the Rabi-rotation of the Bloch sphere and its dilation due to the differential decay create yet another type of continuous transformation that is called the {\it loxodromic} M\"{o}bius transform (in terms of the previously defined complex-variable function $f(z)=(az+b)/(cz+d)$, it is described by $a=\exp\{(i\Omega_R-\gamma_R)t\}$, and $b=c=0$, $d=1$, as before). Under such a transformation, the resulting trajectories on the Bloch (Riemann) sphere form a family of loxodromes (rhumb lines), an example of which is shown in Fig.~\ref{FIG_L2_model}e for the QS evolution in a given arbitrary point in space $\bm{r}_i$. The corresponding trajectories of quantum states on the $(x,y)$--plane are double logarithmic spirals ({\it spira mirabilis}) in the bipolar circular coordinates ($s,\varphi_{LU}$),  pointing away from the point `U' and towards the point `L', with the slope of the spiral depending on the decay to Rabi frequency ratio $\gamma_R/\Omega_R$, providing a circle for $\gamma_R=0$. The rhumb angle with respect to the isocontent parallels is precisely defined by $\tan\alpha = \gamma_R/\Omega_R$ both on the sphere and in real space, due to the conformal mapping. Figure~\ref{FIG_L2_model}f shows the trajectories of a given QS (defined by a fixed Bloch sphere unit vector $\bm{s}_i=(s,\varphi_{LU}^0)$), for two values of the ratio $\gamma_R/\Omega_R$. With time, all states follow this evolution towards the longer surviving LP normal mode (thus the total density shifts to the south of the sphere); at the same time, each point of the real space increases its LP content, so that the isodensity lines ($s=\text{const}$) stereographic projections shift from the region around the point `L' towards the point `U'. Finally, when the vortex cores in all fields coincide in the fixed position of the LP core, the pure UP state acquires zero density.

\section{Vortex velocity and momentum}

\subsection{Core motion and superluminal velocity}

\begin{figure*}[htbp]
  \centering
  \includegraphics[width=1.00\linewidth]{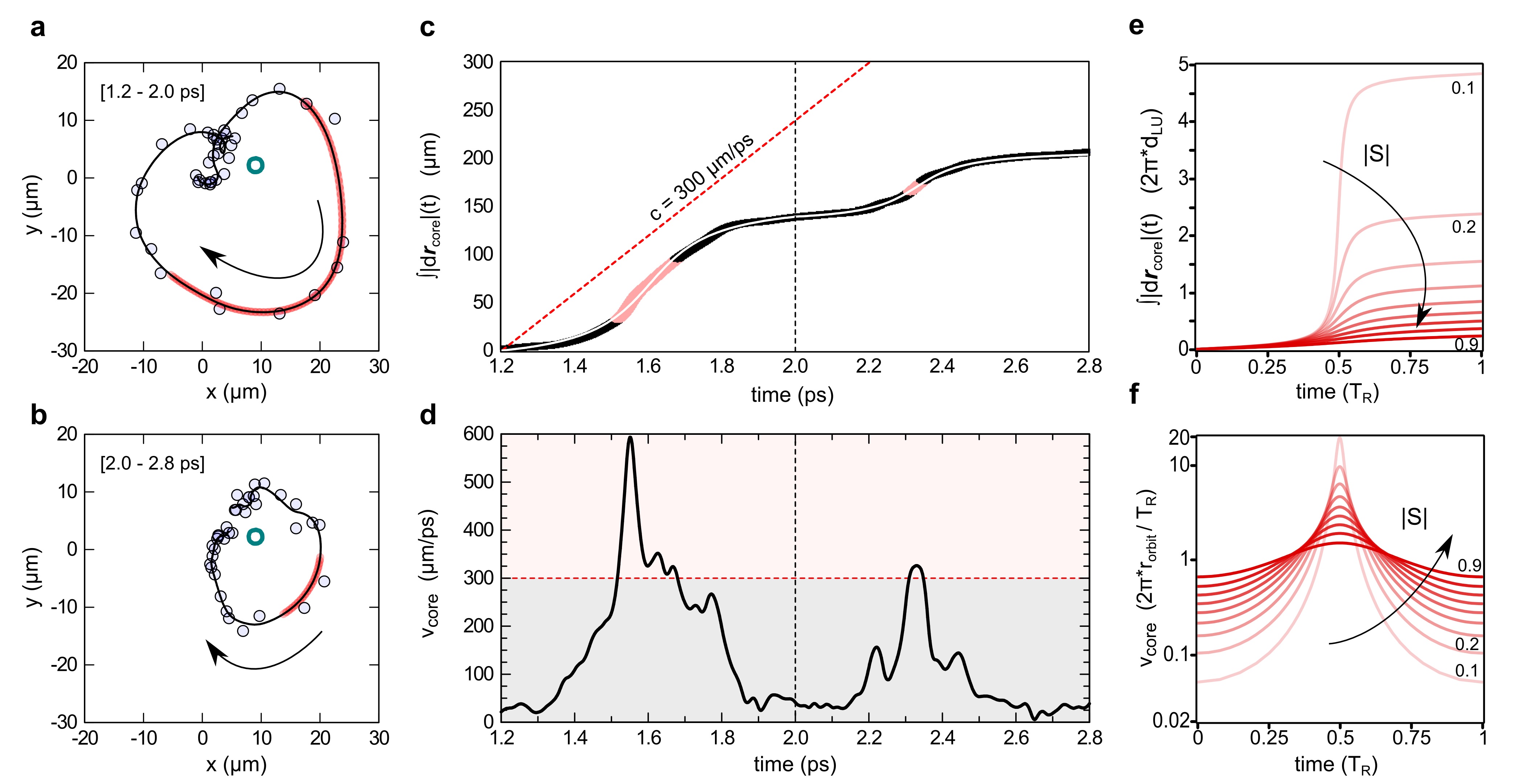}
  \linespread{1.0}\protect\caption{\textbf{Vortex core dynamics.}
  \textbf{a,b}, In-plane vortex core positions (grey open dots) and trajectory (solid line) swept during the first (\textbf{a}) and second (\textbf{b}) Rabi loops. The red shading of the lines shows when the vortex core moves at superluminal speed with $v_\mathrm{core} \geq c =
  300~\mu$m~ps$^{-1}$. The green open dot in the centre maps the position of the LP vortex core (extracted from the experimental data as explained in the Appendix).
  \textbf{c}, Experimental integrated distance swept by the observable photonic vortex core along its trajectory during the first two Rabi cycles. The dashed lines have a slope corresponding to the speed of light $c$, as a reference. The color code is the same as in \textbf{a,b}. The superimposed white solid line results from merging the two separate fittings of the distance by an inverse tangent during each orbital time range (1.2--2.0 and 2.0--2.8~ps).
  \textbf{d}, Velocity of the vortex core during the first two Rabi cycles with the same color code as in \textbf{a}--\textbf{c}.
  \textbf{e}, Model of the integrated distance swept by the core in one Rabi cycle, for different orbits parameterized by the global polaritonic imbalance ($S$ values ranging from 0.1 to 0.9, as marked). The distance is expressed in terms of the distance~$d_{LU}$ between the LP-UP cores (which is constant for a given realization).
  \textbf{f}, Model of the instantaneous velocity of the core, for different global imbalances
  (with same color code as in \textbf{e}). The speed is normalized to the average velocity along a given orbit ($\langle v_\textrm{core}\rangle = 2\pi r_\textrm{orbit}/T_R$), and is expressed in log scale to compress the different aspect ratios in the same panel.  }
\label{FIG_exp_velocity}
\end{figure*}

The average speed of the vortex core along an orbit depends on the orbit size ($\propto r_\textrm{orbit}$), as the oscillation period is fixed by the Rabi coupling. For example, in the
experiments of Fig.~\ref{FIG_exp_rartex_maps}d, the photonic phase singularity is sweeping a closed $xy$ curve with the mean radius of $\SI{20}{\micro\meter}$ during the first Rabi loop, resulting in an average speed $\sim\SI{150}{\micro\meter\per\pico\second}$ which is half the speed of light $c$ in vacuum. Moreover, as noted above, the vortex core speed is not constant: the largest speed is achieved when the core is far from the centre of the spot, in the orbital apex position. A natural question arises regarding the maximum possible speed of rotation. The homeomorphism allows us to easily track the velocity at which the vortex core travels during the dynamics. Since the stereographic projection of the metric formed by the isodensity lines ($s=$~const) and isophase lines ($\varphi_{LU}=$~const) from the sphere to the plane conserves their orthogonality, the instantaneous tangential velocity of such motion   along $s=0$ follows the phase gradient $\bm{\nabla}\varphi_{LU}$ (see the streamlines in Fig.~\ref{FIG_theo_basic}e,~g and the Supplementary Movie SM4~\cite{SMmovies}). Analogously with the usual (single mode) group speed being expressed in terms of the infinitesimal differentials $v_{g} = d\omega/dk$, here one can write for the core velocity ${v}_\textrm{core} = \Delta \omega / |\bm{\Delta k}| = (\omega_U - \omega_L)/|\bm{k}_U-\bm{k}_L| = \Omega_R /|\bm{\nabla}\varphi_{LU}|$. Hence, since the gradient streamlines for a vortex dipole are circles coinciding with the isodensity orbits $s(x,y)=\text{const}$ (in the considered configuration of concentric $LG$ beams of the same size), the vector expression for the core velocity has the form $\bm{v}_\textrm{core} = \frac{\Omega_R}{|\bm{\nabla}\varphi_{LU}|^2}
\bm{\nabla}\varphi_{LU}$.  The gradient $|\bm{\nabla}\varphi_{LU}|$ is larger in the middle of the vortex dipole, hence in the inner region the two moving cores are slower than at the boundary, where they move faster (see the color of the streamlines in Fig.~\ref{FIG_theo_basic}e). From the geometrical point of view, because the rotation of the Bloch sphere corresponds to an inversion (elliptical M\"obius transform), such a drift that is a uniform on the sphere, when projected on the plane, suffers the typical M\"obius stretch of circles, which is the reason for the greatly varying velocities with drastic accelerations and decelerations of the vortex core (or of any chosen QS).

We present in Fig.~\ref{FIG_exp_velocity}a,b the counterpart of Fig.~\ref{FIG_exp_rartex_maps}d but for expanded beams sizes where the vortex core can complete its Rabi cycle with a superluminal maximum speed. Its in-plane $xy$ orbits during the first and second Rabi loops are obtained from its instantaneous positions tracked every 20~fs and indicated by circles, while the lines are mapping the trajectory. The integrated distance swept by the vortex core and the corresponding instantaneous velocities, retrieved from the latter, are shown in Fig.~\ref{FIG_exp_velocity}c,d, respectively.
As one can see, the vortex goes well over the speed of light (twice as fast at its peak velocity) and does so cyclically with intervals where it gets almost at rest, implying drastic accelerations and decelerations. The panel Fig.~\ref{FIG_exp_velocity}e shows the integrated distance swept by the core along a given orbit during the Rabi period as numerically reproduced for different values of the global UP-LP population imbalance parameter $S$. The sigmoid shape of these curves can be approximated by an inverse tangent of time, and the corresponding experimental curve in Fig.~\ref{FIG_exp_velocity}c is fitted with two different inverse tangent dependencies for the first and the second Rabi loops, respectively. The corresponding computed orbital speed of the core $v_\textrm{core}$ normalized to the average speed along an orbit ($\langle v_\textrm{core}\rangle = 2\pi r_\textrm{orbit}/T_R$) is shown in Fig.~\ref{FIG_exp_velocity}f. In both Fig.~\ref{FIG_exp_velocity}e and f, we are using on the horizontal axis the wrapped orbital time along a generic Rabi cycle, where $t=0.5T_R$ corresponds to the rotating vortex reaching the apex position.

While superluminal motion has been previously  discussed~\cite{Jakiel1998,kondakci_optical_2019}, typically in experiments performed in the region of anomalous dispersion of an absorbing material and in tunneling experiments~\cite{Jakiel1998}, or upon proper space-time shaping of a wavepacket~\cite{kondakci_optical_2019}, neither of which however applies to polaritons. Here, this observation follows from the fact that the correct description of a full-Bloch beam is in terms of the full wavefunction defined on its appropriate space, rather than in terms of any single-component field. In fact, we note that one cannot properly speak of a {\it field} when there is no QS that is present in some compact extension of the space. Instead, what one deals with is a projection of the full   wavefunction onto a certain observable basis, which can indeed host superluminal artifacts for any isolated quantum state, which is, in the linear regime of our experiment, more properly seen as a phase singularity. The vortex core is one such particular isolated quantum state (the exciton when observing photons). The Rabi-rotating vortex in the linear regime is thus a real-world implementation of the famous {\it superluminal scissor}. Just as the meeting point of two closing blades can go faster than the speed of light, not being an actual physical object, the core in the projected photon field $\langle x,y|C\rangle\langle C|QS\rangle$ (see Eq.~(\ref{projection})), which arises from the underlying full-Bloch beam, is similarly a contrived meeting point of the polariton Hilbert space eigenstates---the upper and lower polaritons (the counterparts of the scissors upper and lower blades)---being  out-of-phase. When they are in-phase, they similarly contrive to render the excitonic core instead, by projecting $\langle x,y|X\rangle\langle X|QS\rangle$.

\begin{figure}[h]
  \centering
  \includegraphics[width=1\linewidth]{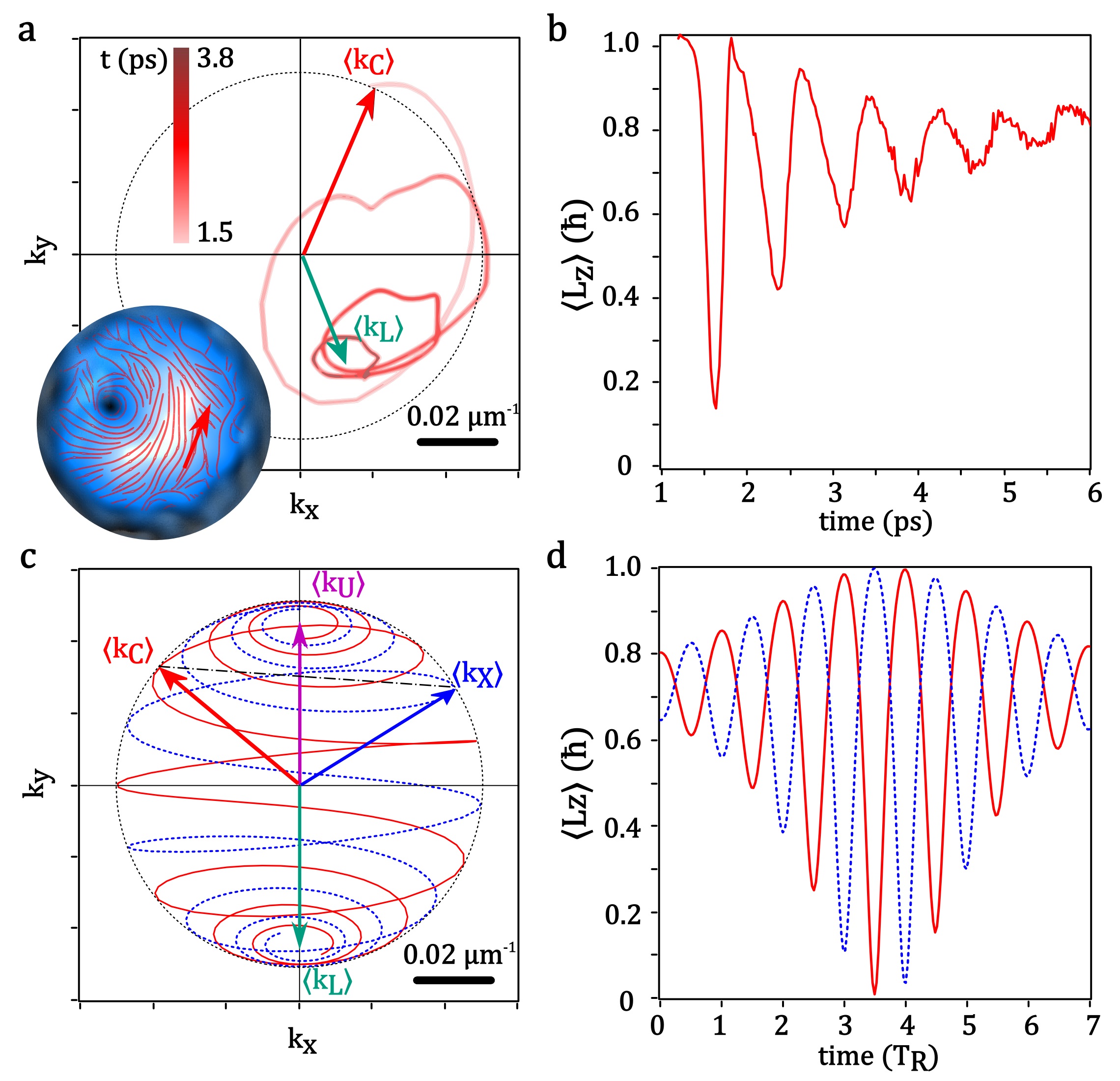}
  \linespread{1.0} \protect\protect\caption{ \textbf{Linear and angular momentum dynamics.}  \textbf{a}, Experimental mean transverse linear momentum $\langle\bm{k}_\mathrm{C}\rangle$ (per particle) in the photon field as a function of time (red arrow and red line, time range 1.5 to 3.8 ps) and in-plane vector momentum of the LP mode $\langle\bm{k}_\mathrm{L}\rangle$ (green vector) extrapolated as the limit vector of the photonic one. The inset shows the density map overlaid with the streamlines of momentum corresponding to the {\it azimuthal} gradient of the phase at $t=1.50$~ps (for the full gradient $k_C=\nabla\varphi$, see the maps in Fig.~\ref{FIG_exp_rartex_maps}d).
  \textbf{b}, Experimental OAM (per particle) with respect to the centre of the initial $LG_{01}$ pulse. Panels \textbf{a},\textbf{b} are relative to the same experimental realization as presented in Fig.~\ref{FIG_exp_rartex_maps}.
  \textbf{c}, Modelisation of the transverse mean linear momentum vectors in the four components, $\langle\bm{k}_{U,L,X,C}\rangle$. The momenta in the normal modes are fixed and pointing upwards and downwards because the vortex cores in these fields are displaced to the left and right, respectively. The total momentum $\langle\bm{k}\rangle$ keeps always vertical, drifting from $\langle\bm{k}_{U}\rangle$ to $\langle\bm{k}_{L}\rangle$ in time due to differential decay. The linear momenta in the bare fields oscillate (red solid / blue dashed for $\langle\bm{k}_{C}\rangle$ / $\langle\bm{k}_{X}\rangle$, respectively), in such a way that their vectorial sum is always instantaneously null in the horizontal direction. Their vertical components are instead oscillating around the shifting vector $\langle\bm{k}\rangle$.
  \textbf{d}, Mean orbital angular momentum in the photon (red solid) and exciton (blue dashed) field with respect to the initial centred axis of symmetry. Both {\bf c},{\bf d} are modeled for a situation of $\pi$-opposite displacement of the UP and LP vortex cores, with the initial and final conditions $S=+0.98$ and $S=-0.98$, respectively, and $\Omega_R/\gamma_R = 10$.  }
\label{FIG_linear_k_map}
\end{figure}

\subsection{Time-varying linear and angular momentum}

A complementary picture to the vortex rotational dynamics is provided by the evolution of the linear and angular momenta. Although both pulses $A$ and $B$ have no mean transverse linear momentum (mTLM) $\langle\bm{k}\rangle=0$, and exactly one unit of angular momentum $l_A=\pm1$ from the pulse~$A$, when the rotational motion of the vortex is triggered, the displacement of the vortices in the LP and UP components from the origin corresponds to a nonzero net linear momentum in each component, which results in the oscillations of the mTLM and the mean OAM of the bare photon and exciton fields, as shown in Fig.~\ref{FIG_linear_k_map}. In Fig.~\ref{FIG_linear_k_map}a, the mTLM vector $\langle\bm{k}_C\rangle$ experimental dynamics is computed from the momentum streamlines (represented in the inset of Fig.~\ref{FIG_linear_k_map}a for $t=\SI{1.5}{\pico\second}$), weighted by the photon density, for successive moments of time in the range $t=1.5$--$\SI{3.8}{\pico\second}$.
The OAM per particle evolution shown in Fig.~\ref{FIG_linear_k_map}b is retrieved from the measured photon complex wavefunction $\psi_C$ via the direct integration $\langle L_z\rangle=i\int\psi_C^*(y\partial_x - x\partial_y) \psi_Cd\bm{r}/\int|\psi_C|^2d\bm{r}$.
While no external transverse forces that would explain the rotation of the centre of mass and the time varying linear and angular momentum are applied, here one deals with a non-constant mass problem, in which each field, photon and exciton, coherently injects particles into its counterpart. The generalized Newton equation $\bm{F}_\text{ext} = 0 = \hbar (m\,d\bm{k}/dt + \bm{k}\,dm/dt)$ holds ({\it i.e.}, inside the microcavity, in the case of no loss or dephasing, the total quantities are conserved).  This is analogous as a concept to the OAM selective evaporation\cite{guo_supersonic_2020} recently used to increase the mean OAM and create a very large vortex in atomic BECs. Here, the coherent transfer implies the momentum exchange between the \changed{fields}, which is responsible for the periodical change of its instantaneous mean value in any of the two subsystems.

The numerical evolution of the $\langle\bm{k}_{C,X}\rangle$ vector in each field is displayed in Fig.~\ref{FIG_linear_k_map}c. The OAM with respect to the original centre of symmetry $\bm{o}$ of the beams can be expressed as $\bm{L}_o = \bm{L}_{core} + \bm{r}_{core} \times \langle\bm{k}\rangle$, where $\bm{L}_{core}$ is the iOAM computed with respect to the moving centre of rotation, {\it i.e.}, the
instantaneous vortex core, which remains constant and quantized, equal to the initial topological charge $l_A$. In contrast, both $\bm{r}_{core}$ and $\langle\bm{k}\rangle$ oscillate, being expected to reach their maxima simultaneously (maximum net linear momentum for maximally displaced vortex core) and almost orthogonal to each other, so that their cross product gives rise to the OAM oscillations. The oscillations calculated with the wavefunctions $\psi_\textrm{C,X}$ through direct averaging
$\int\psi_\textrm{C,X}^*\hat{L}_z\psi_\textrm{C,X} d\bm{r}/\int|\psi_\textrm{C,X}|^2d\bm{r}$ are reported in Fig.~\ref{FIG_linear_k_map}d, and they agree with the above vectorial formula for $\bm{L}_o$ when substituting mean $\langle\bm{k}_{C,X}\rangle$ from Fig.~\ref{FIG_linear_k_map}c. The result can be understood as follows: in the assumption of the maximal ($\pi$) angular displacement of the UP and LP cores, the origin $\bm{o}$ is exactly in the middle of the UP-LP vortex dipole. For this reason, when the $\psi_\textrm{C(X)}$ vortex core passes in between the points `L' and `U' (as shown in the bottom panel of Fig.~\ref{FIG_L2_model}c for the photon core), the total OAM is equal to the iOAM value $\pm1$. The vortex core of the respective counterpart $\psi_\textrm{X(C)}$ at the same moment is at infinity, which means that this field is vortex-free and its OAM is zero. All the other vortex positions result in the oscillations of $\langle L_z\rangle$ between 0 and 1. Note that the data of  Fig.~\ref{FIG_linear_k_map}a,~b was obtained in the experimental realisation of Fig.~\ref{FIG_exp_rartex_maps}, which means starting with the LP mode global imbalance ($S\approx -0.25$), while the modelisation in Fig.~\ref{FIG_linear_k_map}c,~d was performed with the initial condition $S(0)=0.98$ (with decay to Rabi frequency ratio set to $\gamma_R/\Omega_R=0.1$), {\it i.e.}, starting very close to the UP mode, and the $\psi_\textrm{C,X}$ vortex cores running through the whole double spiral from `L' to `U', changing from clockwise to anti-clockwise direction of rotation. Hence the experimental panel (Fig.~\ref{FIG_linear_k_map}b) needs to be compared only to the right half of the numerical panel (Fig.~\ref{FIG_linear_k_map}d).

The reported oscillations represent a general class of time-varying OAM, similar to the recently reported self-torque feature of light~\cite{rego_generation_2019}. Here however the OAM (per particle) is assuming non-integer values in a periodical fashion, rather than changing continuously. We thus demonstrate the experimental creation of OAM fractional values (associated with displaced, off-axis vortex cores) with ultrafast oscillations (associated with the swirling core position) in the light emitted from the polariton system. Furthermore, we likewise evidence the time variations of the TLM vector, whose amplitude and direction both oscillate and swirl due to the same core radial and azimuthal motions. We point out that the associated features of the so-called self-torque (and here also self-force), are as a matter of fact generated inside the device where light is linearly and coherently interacting with matter (excitons): the underlying picture is that of the Rabi-rotation of the full Bloch beam, being projected on the photon field $\psi_\mathrm{C}(x,y,t)$ at the moment when the photon escapes the cavity to  propagate to the CCD camera (in case of Ref.~\cite{rego_generation_2019}, instead, it is a nonlinear phenomenon).  Despite the fact that each photon, once being emitted, is not changing its momentum, and that its motion in the longitudinal propagation direction is at the speed of light, such photonic emission could be used to drive another physical system\cite{clerici_observation_2016} ({\it e.g.}, an atomic BEC or optically trapped nanoparticles) and to study its response to such an ultrafast stimulus. It would also be interesting to devise the entity of angular momentum transfer~\cite{parkin_measurement_2006} to an external system swept by our spiraling vortices.

\section{Conclusions and outlook}

We have observed experimentally a surprising dynamics of vortices in multi-component Rabi-coupled \changed{fields}, and shown how the interplay of Rabi oscillations and vorticity exhibits ultrafast (and even superluminal) spiraling motions of the vortex core with striking accelerations and decelerations. Such a phenomenology on its own can provide a starting point as substantial as Airy beams, Bessel beams, X–waves~\cite{Zamboni2004,Gianfrate2018}, {\it etc.}, for possible applications, in particular, for micromanipulation of very light particles in the time domain, similarly to optical micromanipulation by Airy wavepackets~\cite{airy_m}.
At a more fundamental level, our experiment could also bring forward important developments regarding the topology of complex light or on the interpretation of quantum mechanics, providing an example of a simple and elegant description in an abstract space not immediately accessible to our physical reality, which becomes counter-intuitive and bizarre in the physical space where the observation is performed.

Such ultrafast Rabi-rotating vortices follow from the very peculiar configuration we create, whereby all the quantum states of the polariton Hilbert space are simultaneously present in the system and mapped
throughout the beam in such a way as to be furthermore present only at a single point in space.  We call such a peculiar state a full-Bloch beam. This extends the concept of a full Poincar\'e beam\cite{beckley_full_2010} that similarly realizes all states of polarization in space, but further involving a quantum dynamics (here the Rabi rotation), which results in the peculiar phenomenology we report.
The description in terms of uncoupled and steady upper and lower polaritonic fields overlooks that the two polariton fields are related in such a way as to exhibit all possible quantum states (their superpositions), implying nontrivial relationships of their amplitudes and phases in space and time. Furthermore, by creating such a peculiar state, we realize, on a physical system, the abstract  mathematical notion of a homeomorphism between the real physical $(x,y)$--plane (including infinity) and the polariton Bloch sphere of quantum states. The metric of parallels and meridians on the sphere is conformally linked to the plane metric of bipolar circular coordinates, or Apollonian circles. This allows us to map analytically all the trajectories for all the quantum states, including vortex cores, in any basis of observables, by means of the M\"{o}bius transformation, and track them through the  stereographic projection between the sphere and the plane. Such connections illustrate how physical interpretations in terms of familiar objects, in this case photons, which are being ultimately detected, may be counter-intuitive when involving extreme topological configurations such as the full-Bloch beam: once emitted by the cavity, photons represent a physical field carrying an optical vortex at any time, which, from one snapshot to the next, can even travel superluminally,
but this effect does not originate from a photon field inside the cavity, since (prior to the measurement) no such field exists: at all times, the pure photon state is represented by only one point in space. As a consequence, any physical observable that is based on the dynamics of vortices in polariton fields (or any coupled multi-component fields, in general) should be taken with caution, and
one should turn to the full wavefunction description instead in cases where a full-Bloch beam or any object with similar features is produced. Our results also provide an extremely simple and fundamental illustration of the breakdown of the physical picture which could help understand more complicated cases and serve as a textbook elementary paradigm of emergent objects acquiring their own special rules (such as entanglement arises from a full-wavefunction description broken onto a particles picture). In particular, further extensions of our results include going beyond the homeomorphic mapping with a sphere, for instance by spreading significantly the wavefunction in momentum space so as to overlap various Rabi rotations of the Bloch sphere (higher order Rabi-rotating vortices), which results in the creation of vortices and anti-vortices and thus extends even further the phenomenologies of these pseudo-objects which arise from the breakdown of the usual picture. Also, a moving vortex, interactions, and non-Gaussian states, in particular nonlocality, impart the Rabi-rotating vortex with similar additional phenomenologies.
These are the topics for future works.

\section*{Acknowledgments}
We would like to thank Romuald Houdr\'{e} and Alberto Bramati for the
microcavity sample, Lorenzo Marrucci and Bruno Piccirillo for the
$q$-plate devices, and Carlos S\'{a}nchez Mu\~{n}oz, Anton Nalitov and
Alexander Romanov for fruitful discussions.  We acknowledge  the ``Tecnopolo per la medicina di precisione'' (TecnoMed Puglia) -- Regione Puglia: DGR n.~2117 del 21/11/2018, CUP:~B84I18000540002 and  ``Tecnopolo di Nanotecnologia e Fotonica per la medicina di precisione'' (TECNOMED) -- FISR/MIUR-CNR: delibera CIPE n.~3449 del~7/08/2017, CUP:~B83B17000010001, the Italian MIUR (Ministry of University and Research) PRIN project InPhoPol, the Russian Foundation for Basic Research (Project No.~19--02--00793), the Iran National Science Foundation (INSF Project No. 9709377), and the Australian Research Council Centre of Excellence in Future Low-Energy Electronics Technologies (Project No. CE170100039) for financial support.

\setcounter{equation}{0}
\renewcommand{\theequation}{A\arabic{equation}}




\section*{APPENDIX}

\begin{figure*}[htb]
  \centering
  \includegraphics[width=\linewidth]{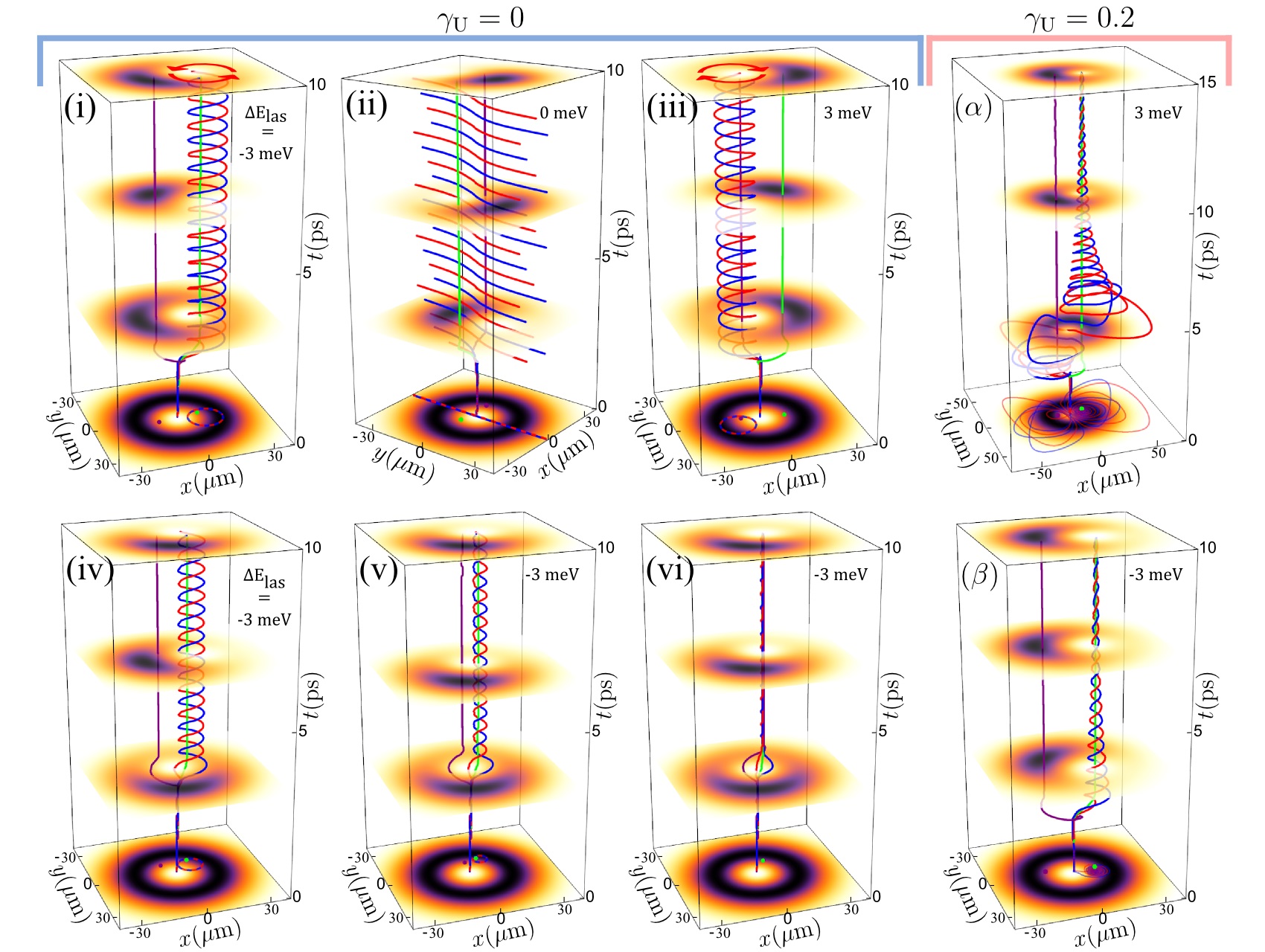}
  \linespread{1.0} \protect\protect\caption{
     \textbf{Trajectories modelisation as a function of different key parameters.}
    \textbf{i-iii}, $xyt$ vortex lines for the photon (red), exciton (blue), UP (purple), LP (green) fields, in the absence of decay ($\gamma_\mathrm{U} = 0$), plotted for $E_\mathrm{las} = \hbar\omega_\mathrm{las}$ detuned from the bottom of the bare fields dispersions by $\Delta E_\textrm{las} = -3$, 0, and \SI{3}{\milli\electronvolt},
    which corresponds to the initial values of the global UP-LP imbalance $S\approx -0.75$, $0$, and $0.75$, respectively.
    \textbf{iv-vi}, Vortex lines at a fixed excitation energy detuning $\Delta E_\textrm{las} = -\SI{3}{\milli\electronvolt}$ ($S(0)\approx -0.75$) for three different time delays $t_{AB}$
    between the two excitation pulses, corresponding to the successive increase of the optical phase shift $\varphi_{AB}$ from the anti-Rabi phase (see \textbf{i} where the UP-LP cores splitting is maximal) up to the Rabi-phase where the UP and LP cores are displaced into the coinciding position (\textbf{vi}).
    $\bm{\alpha},\bm{\beta}$ Simulations with the experimental dissipation rate ($\gamma_\textrm{U} = \SI{0.2}{\pico\second}^{-1}$) included in the cSEs model according to Eq.~(\ref{decay}). In $\bm{\alpha}$, the system is initialized with a larger UP content
    ($S(0)\approx 0.75$), so that the photon and exciton vortices start to rotate around the UP vortex core (purple line). After the population reversal due to the faster UP decay, when $S(t)=0$, the vortices switch to rotating around the LP core (green line), thus repeating the behaviour shown in $\bm{\beta}$, where the excitation energy is closer to the LP mode
    ($S(0)\approx -0.75$). See also the Supplementary Movie SM2~\cite{SMmovies}.  }
\label{FIG_theo_appendix}
\end{figure*}

\subsection{Experimental methods}
The excitation laser is a \SI{80}{\mega\hertz} train of \SI{130}{\femto\second} pulses with a \SI{8}{\nano\meter} bandwidth properly tuned in order to resonantly excite both the LP and UP
branches (with their central $k=0$ states respectively located at 836.2 and \SI{833.2}{\nano\meter} in our sample), which is required to trigger Rabi oscillations.  The modes splitting of \SI{3}{\nano\meter} (\SI{5.4}{\milli\electronvolt}) corresponds to the Rabi period of $T_R =\SI{0.780}{\pico\second}$.
Once imprinted, the polariton vortices are left free to evolve. We operate in a clean area of the sample so to avoid any unintentional effects from the disorder and in the intensity/density regime weak enough in order not to perturb the vortex dynamics by the nonlinearities.
The positional stability of the vortex is indeed observed in the dynamics (in absence of the second/displacing pulse), during the whole LP lifetime (whose decay time is $\tau_{L}\sim \SI{10}{\pico\second}$), with typical Rabi oscillations (quenching with the UP dephasing time $\tau_{U}\sim \SI{2}{\pico\second}$). Standard beam splitters (BS) and $\lambda/4$ plates are used to separate the beams, control the polarization and put them together before sending onto the sample.

In the off-axis digital holography, the resonant emission is let to interfere on an imaging camera with a time-delayed reference beam (detection delay line).  The reference pulse is expanded by passing through a pinhole in order to make it wide and homogeneous. Hence, we apply digital Fourier-space filtering to retrieve both the amplitude and phase of the photon wavefunction. Customized software allows us to monitor and adjust the dynamics in real-time during this operation. The polarization can be simply postselected on the detection side by using $\lambda/4$ or $\lambda/2$ plates and a polarizing beam-splitter.

\subsection{The coupled Schr\"odinger equations model}

We write the two coupled Schr\"odinger (linear) equations according to the standard polaritonic coupled-oscillators model (referred to as cSEs model in the main text):
\begin{equation}
\label{Eq1}
i \partial_t \begin{pmatrix}
\psi_\mathrm{C}  \\
\psi_\mathrm{X}
\end{pmatrix}
=
\begin{pmatrix}
 -\frac{\hbar\nabla^2}{2 m_\mathrm{C}} & \frac{\Omega_R}{2}\\
 \frac{\Omega_R}{2}  & -\frac{\hbar\nabla^2}{2 m_\mathrm{X}}
\end{pmatrix}
\begin{pmatrix}
\psi_\mathrm{C}  \\
\psi_\mathrm{X}
\end{pmatrix}
+\boldsymbol{F}\,,
\end{equation}
where the photonic and excitonic fields $\psi_\mathrm{C,X}(x,y,t)$, with their corresponding parabolic
dispersions with effective masses $m_\mathrm{C,X}$, are coupled via the Rabi coupling term~$\Omega_R$. The excitation scheme is accounted for through the vector $\boldsymbol{F}=(LG_{01}\mathcal{T}_\mathrm{A}+LG_{00}\mathcal{T}_\mathrm{B},0)^T$, acting only on the photonic component.  The $LG_{01}$ and $LG_{00}$ functions describe the vortex and vortex-free spatial shapes of the two pulses.  The functions $\mathcal{T}_\mathrm{A,B}=\mathrm{e}^{-(t-t_\mathrm{A,B})^2/2
  \sigma_t^2} \mathrm{e}^{-i \omega_\mathrm{las}t}$ describe the pulses' time shapes, with the laser
pulses being sent at the different times~$t_{A,B}$ but with the same energy~$\hbar\omega_\mathrm{las}$ and temporal spread $\sigma_t$. The resulting motions of the vortex cores, obtained by tracking the phase singularity points in the 
photon and exciton components $\psi_\mathrm{C,X}$ and the LP and UP fields
$\psi_\textrm{L,U}=(\psi_\textrm{C}\pm\psi_\textrm{X}) / \sqrt{2}$ in space and time, are presented in Fig.~\ref{FIG_theo_appendix}i--vi and the Supplementary Movie SM2~\cite{SMmovies}. The first three panels (i--iii) show the modelisation cases of different detunings of the excitation energy $\hbar\omega_\mathrm{las}$ from of the coinciding bottoms of the bare photon and exciton dispersions. Then, the system is excited closer to
the LP energy, that makes the lower polariton population prevail, {\it i.e.}, $S<0$ (in case of Fig.~\ref{FIG_theo_appendix}i which coincides with Fig.~\ref{FIG_theo_basic}a in the main text, $S\approx-0.75$), the equator of the polariton Bloch sphere (isodensity line  $|\psi_\mathrm{C}|=|\psi_\mathrm{X}|$) is projected on the plane closer to the $|U\rangle$ state, thus the photon and exciton cores start to describe circular orbits around the fixed position of the LP core. In the case of a fully symmetric excitation $S=0$, shown in Fig.~\ref{FIG_theo_appendix}ii, this circle becomes a straight line exactly in the middle between the UP and LP cores positions (corresponding to the limit of a circle with an infinite radius). Finally, Fig.~\ref{FIG_theo_appendix}iii displays the case of
the excitation closer to the UP mode ($S\approx0.75$), in which the C and X cores orbit lies closer to the projection of the $|L\rangle$ state onto the plane, {\it i.e.}, around the UP core. The panels (iv)--(vi) show the modelisation results for a fixed value of $\omega_\textrm{las}$, closer to the LP energy, but for different time delays $t_{AB}$ between the $LG_{01}$ and $LG_{00}$ pulses. One can see that with an increased delay, the angular difference between the displacements of the UP and LP cores changes, which results in the overlap of all cores positions in the case where $t_{AB}$ is equal to an integer number of the Rabi periods $2\pi/\Omega_R$ (see Fig.~\ref{FIG_theo_appendix}vi).

The term $\gamma_\mathrm{U}\gg\gamma_\mathrm{L}$ accounting for the UP lifetime~\cite{Dominici2014} can be introduced in Eqs.~\ref{Eq1}, once written in the polariton basis by diagonalisation:
\begin{equation}\label{decay}
\tilde{H}=
\begin{pmatrix}
  E_\mathrm{L} & 0\\
  0 & E_\mathrm{U}-i \gamma_\mathrm{U}
\end{pmatrix}\,.
\end{equation}
One can come back to the photon/exciton basis by applying the eigenvector matrix $P$ of the original Hamiltonian: $H= P \tilde{H} P^{-1}$. As a result, one can plot, similarly to the lossless case, the $xyt$ vortex lines for the four fields, as show in Fig.~\ref{FIG_theo_appendix}$\alpha$ and~$\beta$. The top panel ($\alpha$, coinciding with Fig.~\ref{FIG_theo_basic}b in the main text) shows the case of the excitation closer to the UP energy and the $\psi_\mathrm{C,X}$ vortices undergoing a double spiral, first
rotating clockwise around the UP core and then anti-clockwise around the LP core. The bottom panel ($\beta$) shows the dynamics after the excitation closer to the LP energy, which corresponds to the experimental case of Fig.~\ref{FIG_exp_rartex_maps} of the main text.

\begin{figure}[ht]
  \centering
  \includegraphics[width=0.97\columnwidth]{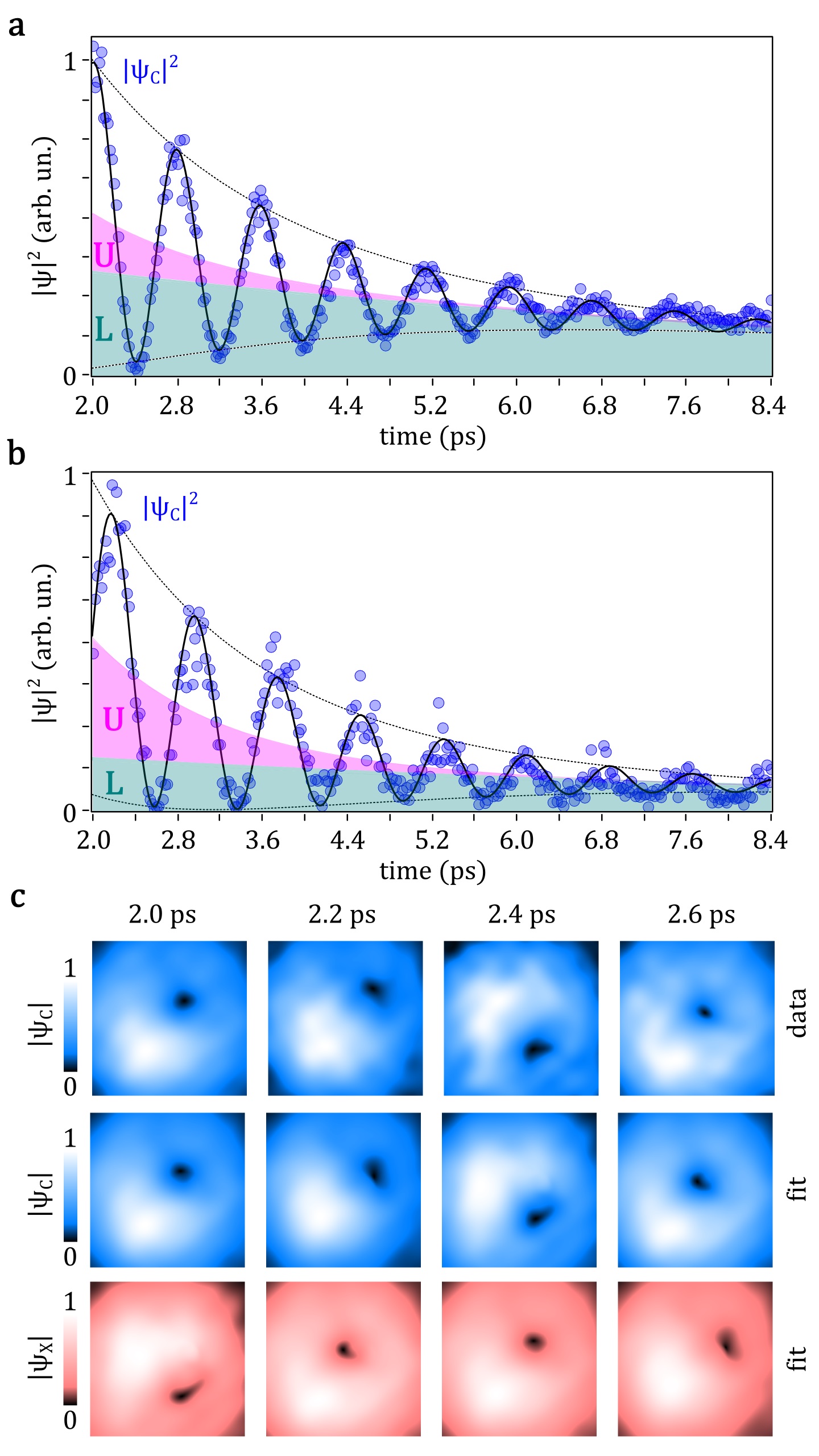}
  \linespread{1.0} \protect\protect\caption{ \textbf{Experimental reconstruction of the
        polariton fields.}
    \textbf{a,b} Experimental oscillations of the photon density (blue
    dots) and their fitting by the theoretical model (black solid line), at two different locations of space (from the same realization of Fig.~\ref{FIG_exp_velocity}a--d and Fig.~\ref{FIG_theo_basic}g) that feature opposite initial local polariton imbalance ($s<0$ and $s>0$, respectively) as well as quadrature initial relative phases in quadrature     ($\varphi_{LU}^0 = 0$, and $-\pi/2$, respectively). The purple and green filled area represent the instantaneous UP and LP densities, respectively.
    \textbf{c}, The original experimental photon density map is shown (first row) together with the reconstruction in time by the theoretical model of both its photon (second row) and exciton (third row) components, for four different times.  }
\label{FIG_interference}
\end{figure}

\subsection{Closed-form solutions and experimental fit}

Since we are in the linear regime, and replacing the sequence of pulses by suitable initial conditions, we can express directly the photon density at each point in space simply in terms of a superposition of the eigenmodes, as follows:
\begin{equation}\label{interference}
|\psi_\textrm{C}|^2 = |\psi_\textrm{L}|^2 + |\psi_\textrm{U}|^2 + 2|\psi_\textrm{L}||\psi_\textrm{U}|\cos(\varphi_{LU}^0+\Omega_R t).
\end{equation}
The UP and LP modes decay can also be included in (\ref{interference}) as $|\psi_\textrm{U,L}(t)| = |\psi_\textrm{U,L}(0)|\exp(-\gamma_\textrm{U,L}t)$. With the fixed experimental values for
$\gamma_\textrm{U}$, $\gamma_\textrm{L}$, and $\Omega_R$, and treating $|\psi_\textrm{L}(\bm{r},0)|$, $|\psi_\textrm{U}(\bm{r},0)|$, and $\varphi_{LU}^0(\bm{r})$ as fitting parameters, we can fit in time and at each point $(x,y)$ of the spot the experimental
photon wavefunction $\psi_\textrm{C}$, with the simple Rabi dynamics  Eq.~(\ref{interference}). This provides the density profiles of the UP and LP polariton fields, as well as their relative phase spatial profile presented in Fig.~\ref{FIG_theo_basic}g. In a similar fashion, the position of the LP core shown in Fig.~\ref{FIG_exp_velocity}a and~b, is calculated from the experimental data by the fitting of
Eq.~(\ref{interference}). The third panel in Fig.~\ref{FIG_theo_basic}g is obtained by inserting the fitted $|\psi_\textrm{L}|$ and $|\psi_\textrm{U}|$ at
$t=\SI{2.0}{\pico\second}$ into the definition for the local UP-LP imbalance parameter:
$s=(|\psi_\textrm{U}|^2-|\psi_\textrm{L}|^2)/(|\psi_\textrm{U}|^2+|\psi_\textrm{L}|^2)$. An example of time fitting (black solid lines) the photon density (blue dots) according
to Eq.~\ref{interference} is shown in Fig.~\ref{FIG_interference}a,~b at different spatial positions that feature different initial local imbalance $s(0)$ ($s(0)<0$ and $s(0)>0$, respectively) and different $\varphi_{LU}^0$ ($\varphi_{LU}^0 = 0$ and $-\pi/2$,
respectively). The fitting is performed along eight Rabi periods in the time range $2.0$ to $\SI{8.4}{\pico\second}$, which is starting from the second Rabi cycle, {\it i.e.}, after the setting sequence of both pulses $A$ and $B$. The comparison of the fitting dynamics with the experimental data are shown in Fig.~\ref{FIG_interference}c, for four times into the second Rabi cycle.

\subsection*{Captions to Supplementary Movies}

\textbf{Movie SM1}:
Experimental vortex dynamics as in Fig. 1. Photonic amplitude and phase in a $100\times100~\mu$m area, with 20 fs time step. The Gaussian beam $B$ is arriving at around $t = 1.6$~ps. The vortex
core position is marked with a yellow dot in the amplitude map.\\

\textbf{Movie SM2}:
Rabi-rotating vortex dynamics simulated by the cSEs model associated to different parameters, and arranged in the same order as in Fig. 7. The $xyt$ vortex lines for each subfield $\psi_\mathrm{C,X,U,L}$ are shown (red, blue, purple and green line for the photon, the exciton, the upper and the lower mode, respectively). The $xy$ photonic density maps are shown at $t = 0$
and current time planes of each graph.\\

\textbf{Movie SM3}:
The interference model showing the maps and dynamics of four relevant quantities, namely the local imbalance $s(\bm{r},t)$ and the relative phase $\varphi_{LU}(\bm{r},t)$ on the top row
(blue-green map and black and white map, as in Fig.~3d,e), and the photon and exciton densities $|\psi_\mathrm{C,X}(\bm{r},t)|$ (blue, red) in the second row. The fixed blue and green dots mark the fixed in space UP and LP cores, respectively. The drifting white and black circles overlapped to the photon and exciton densities represent the $s=0$ isocontent line and the $\varphi_{LU} =0~\cup~\varphi_{LU} = \pi$ isophase lines, respectively. Their crossing points mark the moving photon and exciton cores (dark spots in the respective fields). The global polariton content is changing from $S= 0.93$ to $S= -0.9$.\\

\textbf{Movie SM4}:
The interference model showing the evolution of the total density map $|\Psi_\mathrm{total}|^2
= |\psi_\mathrm{U}|^2 +|\psi_\mathrm{L}|^2$ on the left panel.
The overlapped bipolar coordinates map the Bloch quantum state, {\it i.e.}, the polariton
isocontent $s=\text{const}$ (white circles) and isophase $\varphi_{LU}=\text{const}$ (black
circles) lines, as in Fig.~4a. The panel on the right shows the photon density in time, overlapped with a few specific $s=0$ isocontent lines (white) and the $\varphi_{LU} =0~\cup~\varphi_{LU} = \pi$ circles (black). The color streamlines represent the
vector velocity field of the core (and of any of the other quantum states) changing in time,
retrieved from the relative phase gradient $\bm{v}_\textrm{core} =
\frac{\Omega_R}{|\bm{\nabla}\varphi_{LU}|^2} \bm{\nabla}\varphi_{LU}$, same as in
Fig.~3e.

\def\bibsection{\section*{\refname}}

\end{document}